\documentclass[runningheads]{svmult}
 
\usepackage{makeidx}   
\usepackage{graphicx}  
\usepackage{subeqnar}  
\usepackage{multicol}  
\usepackage{cropmark} 

\def\Dl{D_{\rm L}} \def\Ds{D_{\rm S}} \def\Dls{D_{\rm LS}}
  
\def\etal{{\it et al.}}
\def\eg{{\it e.g.}}

\def\Om{{$\Omega_m$}}
\def\arcm{{$^\prime\,$}}
\def\arcs{{$^{\prime\prime}\,$}}

\begin{document}
\title*{Weak Lensing}


\titlerunning{Weak Lensing}

\author{David Wittman\inst{1}}


\authorrunning{Wittman}

\institute{Bell Laboratories, Lucent Technologies, Room 1E-414,\\
700 Mountain Avenue, Murray Hill, NJ 07974, USA}

\maketitle

\begin{abstract}
In the preceding chapters, the effects of lensing were so strong as to
leave an unmistakable imprint on a specific source, allowing a
detailed treatment.  However, only the densest regions of the universe
are able to provide such a spectacular lensing effect.  To study more
representative regions of the universe, we must examine large numbers
of sources statistically.  This is the domain of weak lensing.
\end{abstract}
\index{weak lensing}

\section{Introduction}
 
\subsection{Motivation}
 
Weak lensing enables the direct study of mass in the universe.
Lensing, weak or strong, provides a more direct probe of mass than
other methods which rely on astrophysical assumptions (\eg\
hydrostatic equilibrium in a galaxy cluster) or proxies (\eg\ the
galaxy distribution), and can potentially access a more
redshift-independent sample of structures than can methods which
depend on emitted light with its $r^{-2}$ falloff.  But strong lensing
can be applied only to the centers of very dense mass concentrations.
Weak lensing, in contrast, can be applied to the vast majority of the
universe.  It provides a direct probe of most areas of already-known
mass concentrations, and a way to discover and study new mass
concentrations which could potentially be dark. \index{lens!dark} 
With sources covering
a broad redshift range, it also has the potential to probe
structure along the line of sight.

Specifically, we might expect weak lensing to answer these questions:
\begin{itemize}
\item Where are the overdensities in the universe ?
\item Are they associated with clusters and groups of galaxies ?  Does
light trace mass in these systems ? \index{galaxy!cluster} 
\index{galaxy!groups}
\item How much do these systems contribute to \Om, 
the mean density of
matter in the universe ?
\item What is their mass function and how does that function evolve
with redshift~? What does that imply for the dark energy equation of state ?
\item What are the structures on larger scales (walls, voids,
filaments) ?
\item Is this structure comparable to that seen in cosmological
simulations?  Which cosmology matches best ?
\item What is the nature of dark matter ? \index{dark matter}
\item Can observations of lensing put any constraints on alternative
theories of gravity ?
\end{itemize}
 
Until recently, deep imaging on the scale required to answer the above
questions with weak lensing was simply impractical.  The development
of large mosaics of CCDs has expanded the field greatly.  The large
data volume leads to ever-decreasing statistical errors, which means
that very close attention must be paid to systematic errors and
calibration issues.  Weak lensing results must be carefully
scrutinized and compared with those of other approaches with this in
mind.

We start with a review of the basic concepts, the limits of weak
lensing, and observational hurdles, and then address the above
astrophysical questions.
 
\subsection{Basics}

\begin{figure}
\centerline{\resizebox{2.4in}{!}{\includegraphics{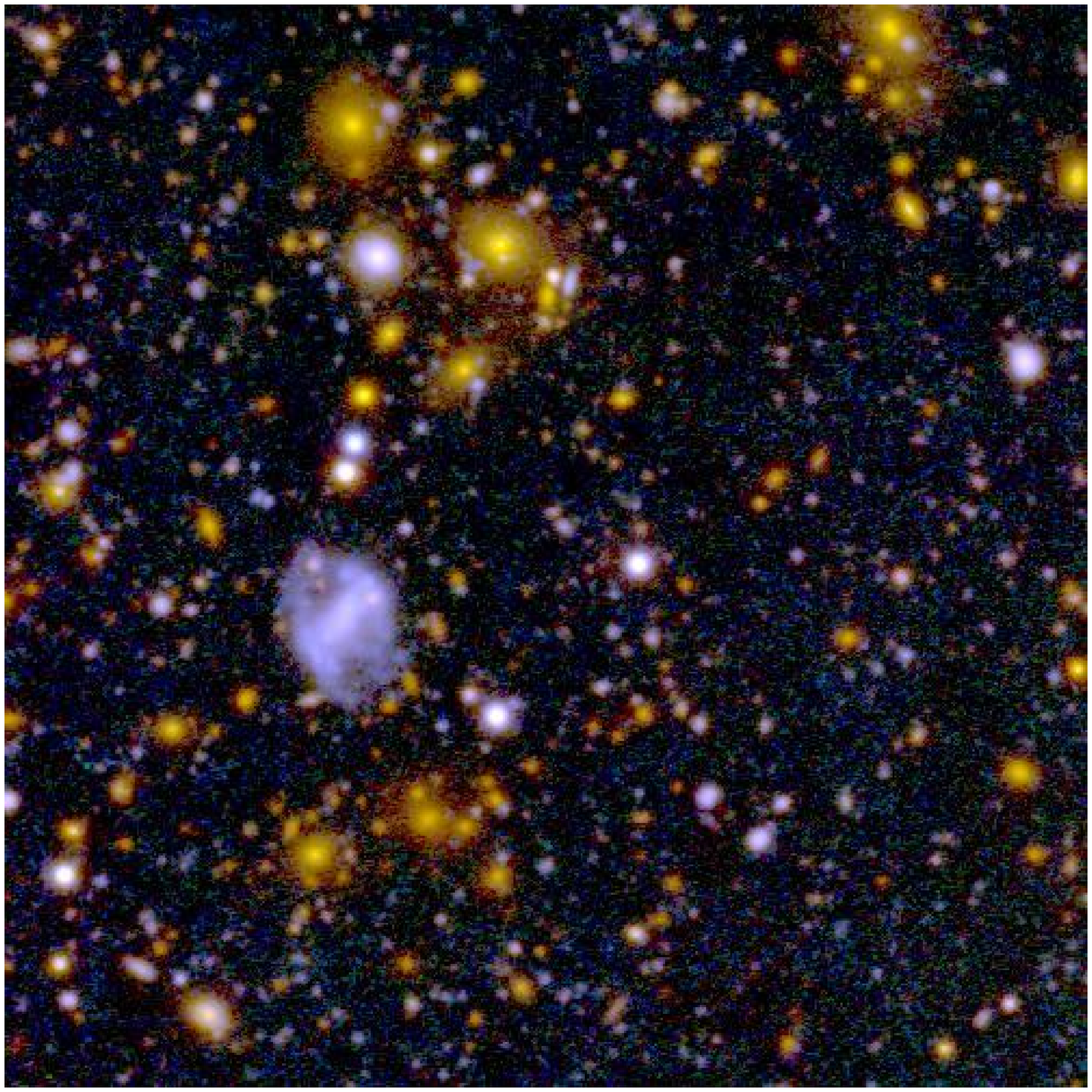}}
\resizebox{2.4in}{!}{\includegraphics{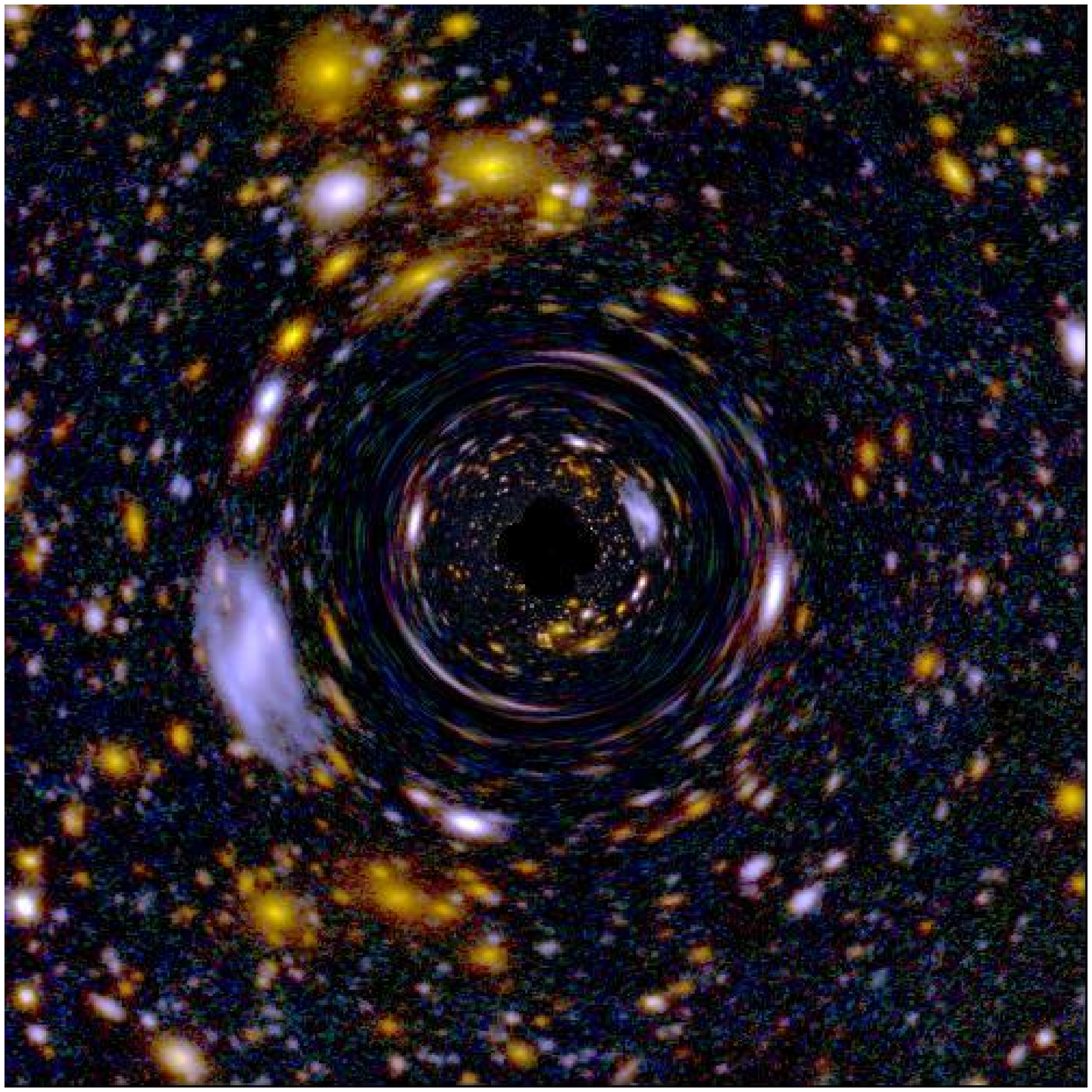}}}
\caption{Simulated effects of a lens: source plane (left) and image
plane (right).  Most regions of the lens can be probed only with weak
lensing.  Real sources are not in a plane, but this does not
dramatically affect the appearance. Real lenses, such as galaxy clusters,
would obscure much of the strong-lensing region.}
\label{fig-demo}
\end{figure}

The transition from strong to weak lensing can be seen at a glance in
the simulation shown in Figure~\ref{fig-demo}.  Over most of the
field, no one galaxy is obviously lensed, yet the galaxies
have a slight tendency to be oriented tangentially to the lens.
We seek to exploit this effect to derive information about the lens,
and perhaps about the weakly lensed sources as well.

We start with the {\it inverse magnification matrix} (see also Chapter 
on quasar lensing)
\begin{equation}
M^{-1} = (1-\kappa)
\left(\begin{array}{cc} 
1 & 0\\
0 & 1\\
\end{array} \right) + \gamma 
\left(\begin{array}{cc} 
\cos 2\phi & \sin 2\phi \\
\sin 2\phi  &  -\cos 2\phi \\
\end{array} \right),
\end{equation}
so called because it describes the change in source coordinates for an
infinitesmal change in image coordinates, the inverse of the
transformation undergone by the sources.  This is Equation 16 of the
Quasar Lensing chapter, which derives $M^{-1}$ and defines the
quantities within.  We repeat here that the {\it convergence} $\kappa$
represents an isotropic magnification, and the {\it shear} $\gamma$
\index{convergence} \index{shear} 
represents a stretching in the direction $\phi$.  They are both
related to physical properties of the lens as linear combinations of
derivatives of the deflection angle.  However, $\kappa$ can be
interpreted very simply as the projected mass density $\Sigma$ 
\index{density!surface} divided
by the critical density \index{density!critical} 
$\Sigma_{\rm crit}$, while $\gamma$ has no
such straightforward interpretation.  In fact, $\gamma$ is {\it
nonlocal}: its value at a given position on the sky depends on the
mass distribution \index{mass!distribution} 
everywhere, not simply at that position.  We will
see this fact rear its ugly head in several places throughout this
chapter.  Shear is often written as a vector $\gamma_i = (\gamma \cos
2\phi,\gamma \sin 2\phi)$ or more succinctly as a complex quantity
$\gamma e^{i2\phi}$.  

Without multiple images of a source (as in the strong lensing case),
we must have some independent knowledge of the sources if we are to
measure magnification or shear. \index{magnification} 
For example, if one source were a
standard candle or ruler, the apparent magnitude or size of its image
would immediately yield the magnification at that point.  Of course,
standard candles or rulers occur only in very special
cases \cite{Blakeslee}, so in practice we must analyze source {\it
distributions}. \index{source!distribution} 
We no longer get much information from a single
source, and thus lose resolution; this is the tradeoff we must make
for probing regions with weak tidal fields.  

One source distribution that could be used in this way is $n(m)$, the
number of galaxies as a function of apparent magnitude.  In practice,
this is difficult, because the measured slope of this distribution
does not differ greatly from the critical slope at which equal numbers
of galaxies are magnified into and out of a given magnitude bin, with
no detectable change ($n \propto m^{0.4}$).  There is enough
difference to make some headway, but we would prefer to measure
departures from zero rather than small changes in a large quantity.

\begin{figure}
\centerline{\resizebox{2.4in}{!}{\rotatebox{-90}{\includegraphics{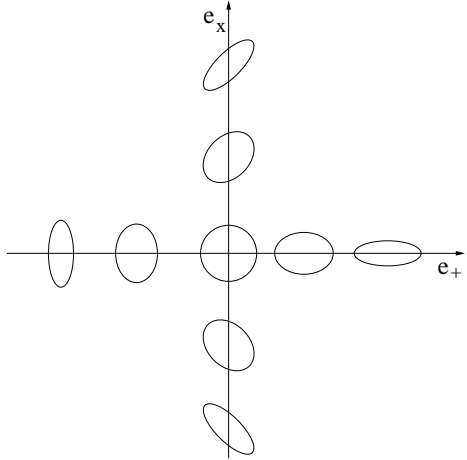}}}}
\caption{A sequence of ellipses with various amounts of each
ellipticity component.  Inspired by the appearance of these ellipses,
the two components \index{source!ellipticity} 
are often labeled $e_+$ and $e_\times$.}
\label{fig-ellip}
\end{figure}

The distribution of galaxy {\it shapes}, \index{galaxy!shape} 
properly defined, does allow
us to measure departures from zero.  Approximate each source as an
ellipse with position angle $\phi$ \index{source!position angle} 
and (scalar) ellipticity \index{source!ellipticity} $\epsilon
= {a^2 - b^2 \over a^2 + b^2}$, where $a$ and $b$ are the semimajor
and semiminor axes.  Define a {\it vector
ellipticity} \index{ellipticity!vector} 
$e_i = (\epsilon \cos 2\phi,\epsilon \sin 2\phi)$, or
equivalently a {\it complex ellipticity} 
\index{ellipticity!complex} \index{polarization} 
$\epsilon e^{i2\phi}$ (also
called {\it polarization}).  This encodes the position angle and
scalar ellipticity into two quantities which are comparable to each
other; the dependence on $2\phi$ indicates invariance under rotation
by 180$^\circ$.  Figure~\ref{fig-ellip} gives a visual impression of
ellipses in this space.

We can now quantify the visual impression of Figure~\ref{fig-demo}.
In the absence of lensing, as in the left panel, galaxies are randomly
oriented: The observed distribution of $e_i$ is \index{source!distribution} 
roughly Gaussian with
zero mean and an rms of $\sigma_e \sim 0.3$.
In the presence of lensing, as in the right panel, this distribution
is no longer centered on zero, as long as we consider an
appropriately-sized patch of sky.  In fact, we will {\it assume} that
any departure from zero mean must be due to lensing.  We will examine
the limits of this assumption in some detail later, but for now let us
accept that on large enough scales, the cosmological principle demands
it, and as a practical matter, we average over sources at a wide range
of redshifts, which are too far apart physically to influence each
other's alignment.

The effect of the magnification matrix on the complex ellipticity can
be computed if $M$ is constant over a source. \index{magnification} 
This is obviously not
valid for very large sources or those near caustics, \index{caustics}
but it is valid
for the vast majority of the sky and for typical sources with sizes of
a few arcseconds.  The result is that $\epsilon^I = \epsilon^S +
{\gamma \over 1-\kappa}$, where superscripts indicate image and source
planes  \cite{Blandford91}.  We don't know any of these quantities for a
single source, but we do know (or assume for now) that $\langle
\epsilon^S \rangle = 0$, where brackets indicate averaging over many
sources. Hence
\begin{equation}  
\label{eq-shear}
\langle \epsilon^I \rangle =  \langle {\gamma \over 1-\kappa}\rangle. 
\end{equation}
The quantity on the right is called the {\it reduced shear} $g$. 
\index{shear!reduced}
A second approximation we can often make is that $\kappa \ll 1$, so that 
$  \langle \epsilon^I \rangle = \langle \gamma \rangle$.  This is called
the {\it weak lensing limit}.  \index{weak lensing!limit}

The fundamental limit to the accuracy with which we can measure
$\gamma$ in the weak lensing limit is {\it shape noise}, 
\index{shape noise} or the width
of the source ellipticity distribution $\sigma_e \sim 0.3$.  Averaging
over $n$ sources should decrease the uncertainty to ${\sigma_e \over
\sqrt{n}}$, but $n$ is limited by the depth of the observations and
the area over which we are willing to average $\gamma$; these
tradeoffs are discussed below.  Also note that knowledge of the shear
alone is not strictly enough to infer mass distributions because of
the {\it mass sheet degeneracy}  \cite{FGS85,SS95}, 
\index{degeneracy!mass sheet} introduced in a
different context in the Quasar Lensing chapter.  This degeneracy
arises because a uniform sheet of mass induces only magnification, 
\index{magnification} not
shear. \index{shear} 
Because the equations are linear, we could therefore add or
subtract a mass sheet without affecting the shear.  In practice, we
can still answer many questions with shear alone, as discussed below.

\subsection{Cosmology dependence}
\index{cosmology}

\begin{figure}
\centerline{\resizebox{3in}{!}{\includegraphics{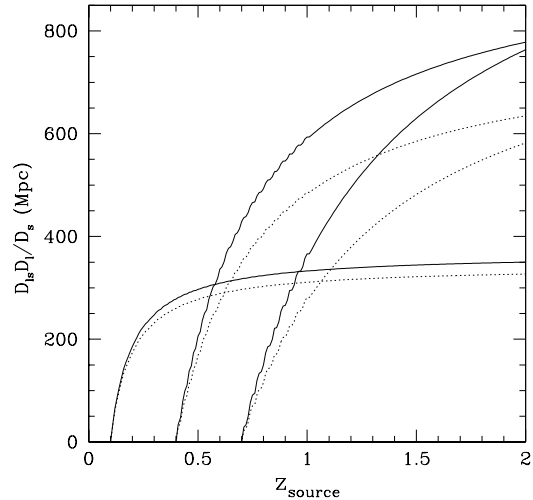}}}
\caption{${{\Dls \Dl} \over \Ds}$ as a function of source redshift,
for several lens redshifts (indicated by the intersections of the
curves with the horizontal axis) and several cosmologies.  The
cosmologies are $\Lambda$-dominated (solid lines, $H_0= 70$ km
s$^{-1}$ Mpc$^{-1}$, $\Omega_m = 0.3$, $\Omega_\Lambda = 0.7$) 
and open 
(dashed lines, $H_0= 70$ km s$^{-1}$ Mpc$^{-1}$, $\Omega_m =
0.4$, $\Omega_\Lambda = 0$).  Each solid line is higher than its
dashed counterpart, reflecting the larger size of the $\Lambda$-dominated
universe.  Although this quantity appears to be a sensitive test of
the cosmology, it is degenerate with the lens mass.}
\label{fig-dratio}
\end{figure}

Both convergence and shear \index{convergence} \index{shear} 
scale as the combination of angular
diameter \index{distances!angular diameter} 
distances ${\Dls \Dl \over \Ds}$, or as the {\it distance
ratio} ${\Dls \over \Ds}$ for a given lens.  (Recall from the
Quasar Lensing chapter that $\Dls$, $\Dl$, and $\Ds$ are the angular
diameter distances from lens to source, observer to lens, and observer
to source, respectively.  Note that $\Ds \ne \Dl + \Dls$; see
 \cite{Hogg} for a quick review and  \cite{Peebles} for a thorough
treatment of distance measures in cosmology).  This
cosmology-dependent quantity is plotted as a function of source
redshift in Figure~\ref{fig-dratio} for several lens redshifts and two
different cosmologies.  In principle, this could be used to constrain
the cosmology if source redshifts \index{source!redshift} 
are known, and if the lens mass is
known independently (the effects of a larger lens mass and a
larger universe are degenerate).  But this remains an unused
cosmological test because lens parameters and source redshifts are
usually poorly known.  Usually, a cosmology is assumed and lens
parameters are estimated using any available knowledge of source
redshifts. Less often, a well-characterized lens is used to explore
the source redshift distribution.  However, source redshift
distributions are usually quite broad, and weak lensing can only be
used to estimate the {\it mean} distance ratio to a group of sources,
which is not same as the distance ratio corresponding to the mean
redshift.  Section~\ref{sec-srd} deals with ways of estimating the
mean distance ratio or otherwise accounting for a broad source
redshift distribution.

\begin{figure}
\centerline{\resizebox{3in}{!}{\includegraphics{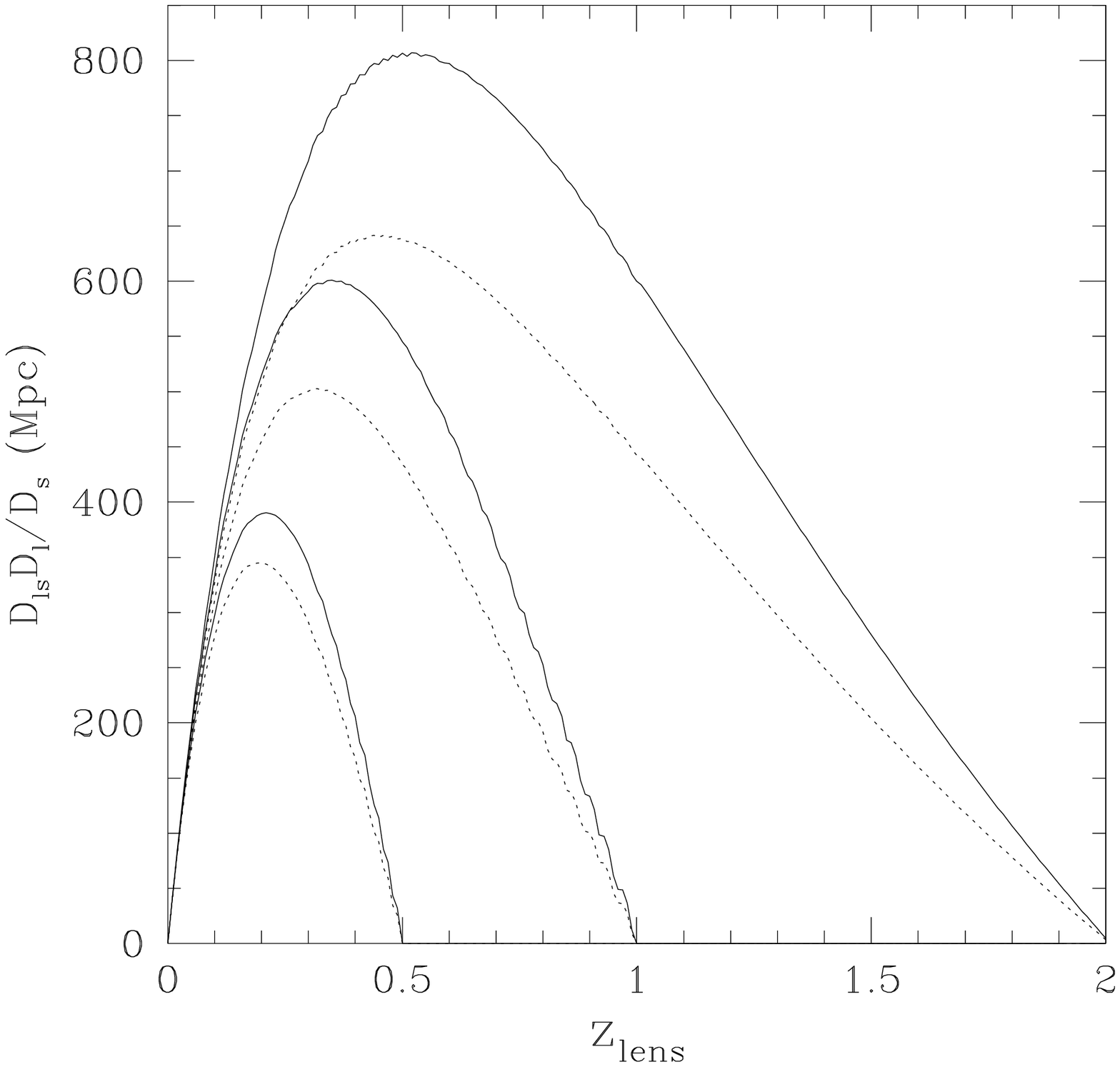}}}
\caption{Same as for Figure~\ref{fig-dratio}, but as a function of
lens redshift, \index{lens!redshift} 
for several values of source redshift \index{source!redshift} 
(which correspond
to the right-hand end of each curve).  The lensing efficiency is a
very broad function, making it difficult to separate unrelated
structures along the line of sight. }
\label{fig-efficiency}
\end{figure}

Another way of viewing the same information is to fix the source
redshift and plot this ratio as a function of lens redshift
(Figure~\ref{fig-efficiency}).  This reveals the relative importance
of different structures along the line of sight and is often called
the {\it lensing kernel} or {\it lensing efficiency}. 
\index{lensing!efficiency}\index{lensing!kernel}

\subsection{Applicability of weak lensing}

As with all astrophysical tools, we must be aware of the limitations
of weak lensing before plunging into results.  They include the weak
lensing approximation itself; \index{weak lensing!approximation} 
mass sheet degeneracy if only shear is
used; poor angular resolution because of its statistical nature;
source redshift difficulties; and possible departures from the
assumption of randomly oriented sources.  We now examine these limits
and ways of dealing with them.

\subsubsection{Weak lensing approximation}

The approximations that $M$ is constant over each source and that
$\kappa \ll 1$ cannot be applied when dealing with the centers of
massive clusters and galaxies. \index{galaxy!cluster} 
Of course, analysis of such regions is
not lacking---it is the topic of most of this book.  Here we merely
wish to mention work that has been done on combining weak and strong
lensing information  \cite{ASW1998}.  We also note that, where only the
second approximation fails, Equation (\ref{eq-shear}) can be solved
iteratively for $\kappa$.

\subsubsection{Mass sheet degeneracy}
\index{degeneracy!mass sheet}

Mass sheet degeneracy was a serious concern when fields of view were
small and lens mass distributions extended well beyond the edges.
Modern imagers now deliver fields of view $\sim 0.5^\circ$ on a side
($>3$ Mpc radius for any lens at $z>0.15$), so this concern has
diminished.  The degeneracy may also be broken by adding magnification
information, which may come from strong lensing, or from a
method called the depletion curve. \index{depletion curve}

\begin{figure}
\centerline{\resizebox{2.4in}{!}{\includegraphics{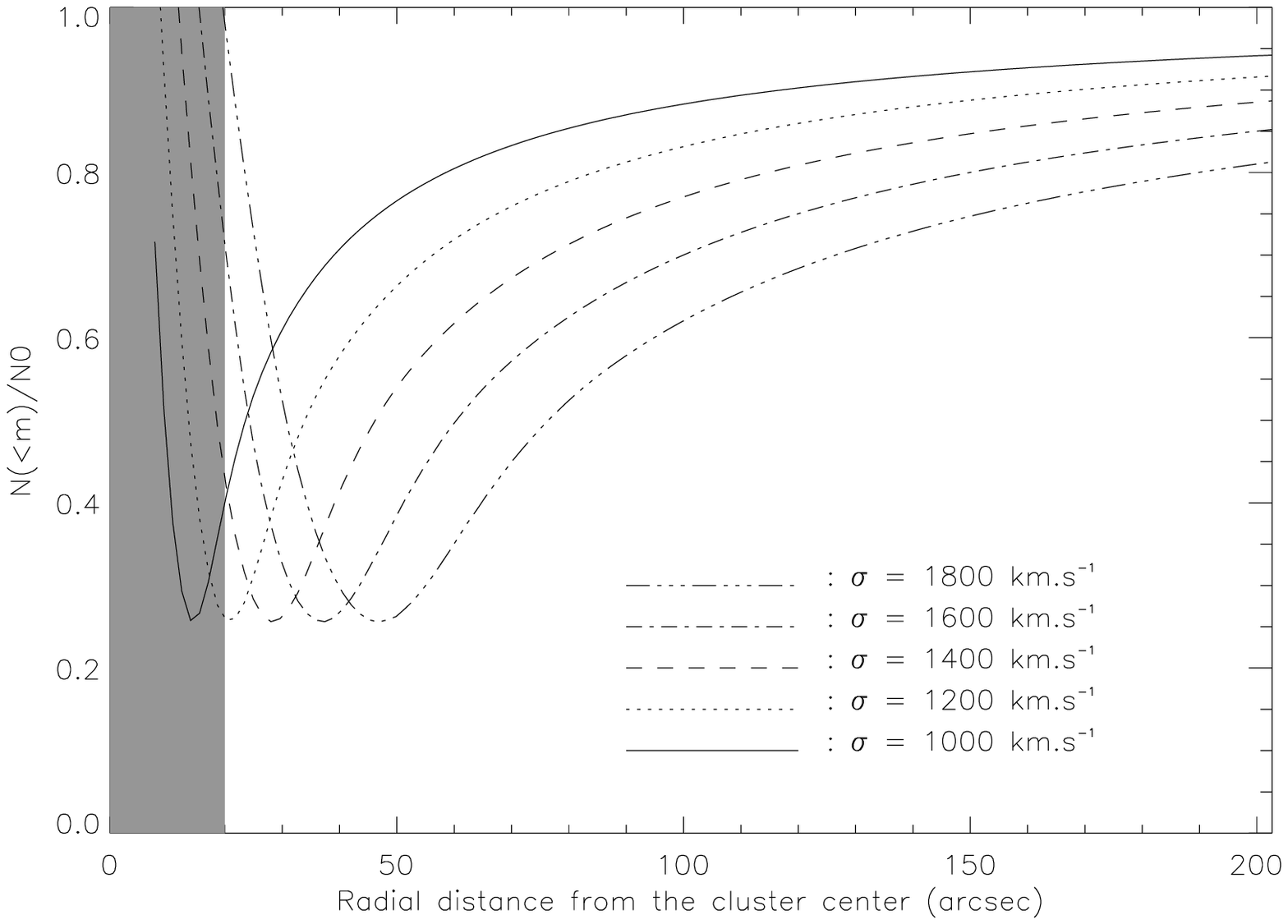}}
\resizebox{2.4in}{!}{\includegraphics{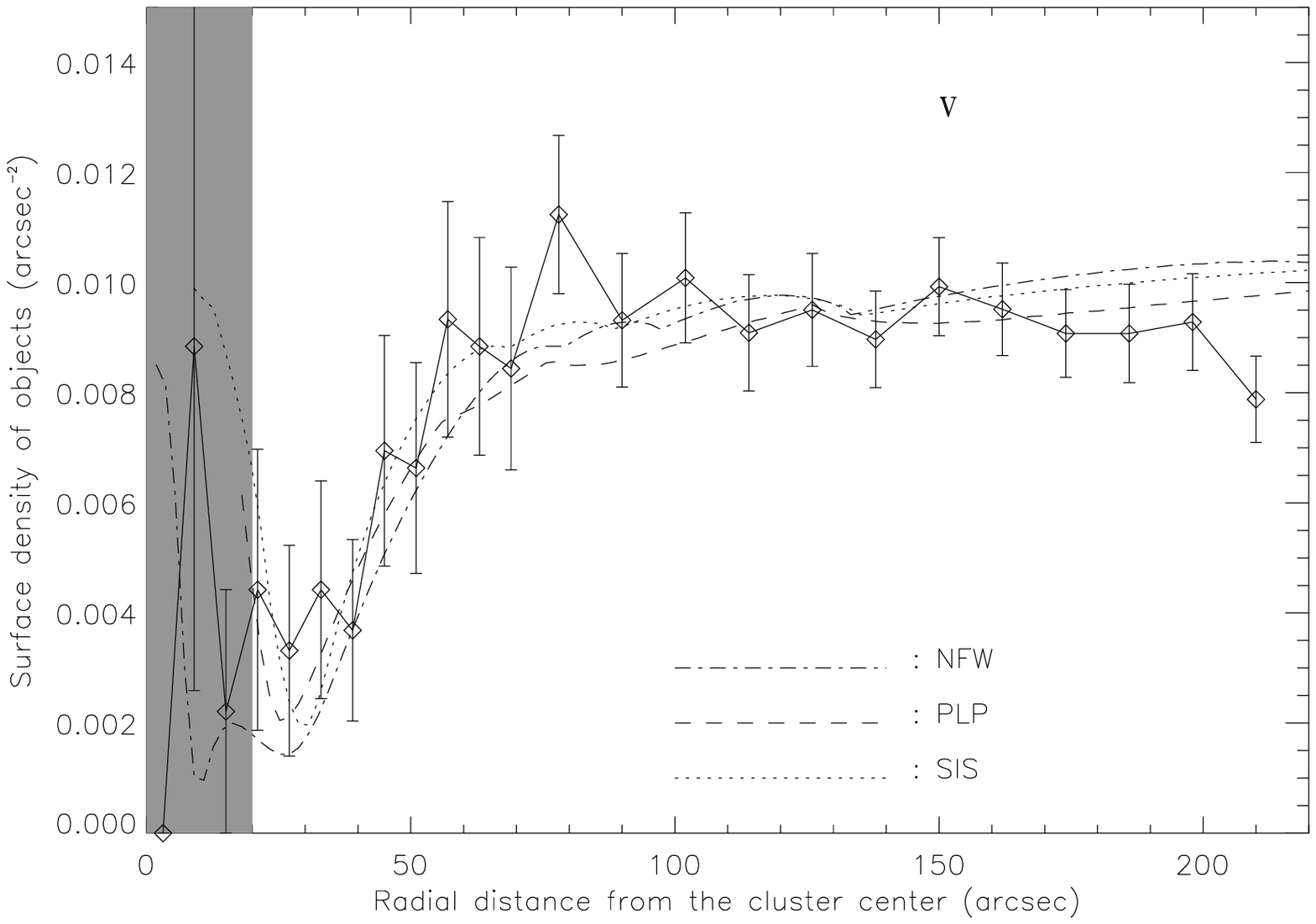}}
}
\caption{Left: theoretical depletion \index{depletion curve} 
curves for a variety of lens
velocity dispersions (lens mass $\propto \sigma_v^2$).  Right:
depletion curve observed for MS1008-1224 in V band.  From
 \cite{MS2000}.}
\label{fig-depletion}
\end{figure}

Magnification \index{magnification} 
imposes two effects of opposite sign on the areal
density of sources. \index{source!density} 
Galaxies fainter than the detection limit (or any
chosen brightness threshold) are amplified above the threshold,
increasing the density of sources, but at the same time the angular
separation between galaxies is stretched, decreasing the density of
sources.  The net effect depends on the slope of $n(m)$, the
(unlensed) galaxy counts as a function of magnitude.  A logarithmic
slope less than 0.4 (usually the case at visible wavelengths, but
barely) will not provide enough ``new'' sources to overcome the
dilution effect, so the source density decreases as $\kappa$ increases
toward the center of a cluster.  This {\it depletion curve} reveals
lens parameters, as shown in Figure~\ref{fig-depletion}  \cite{MS2000}.
Despite the name, the method need not be restricted to one-dimensional
information \cite{BTP95}; \cite{MS2000} includes a lens ellipticity 
\index{lens!ellipticity} \index{lens!position angle} and
position angle estimate based on a crude depletion map.  In practice,
measuring magnification is quite difficult, because the slope of
$n(m)$ is perilously close to 0.4, and there are few published
depletion curve measurements \cite{MS2000,D2001}. For the remainder of
this work, we shall concentrate on algorithms and results using shear,
not magnification.

\subsubsection{Angular resolution}
\index{source!density}

The angular resolution of weak lensing is limited by the areal density
of sources.  With a shape noise \index{shape noise} 
of $\sigma_e \sim 0.3$ and $\sqrt{n}$
statistics, a shear measurement accurate to $p$ percent requires $\sim
1000 p^{-2}$ sources.  Angular resolution is then set by the area of
sky over which these sources are scattered.  This in turn depends on
the depth and wavelength of the observations; in $R$ there is one
source per square arcminute in a one-magnitude wide bin at $R \sim
21.4$, increasing by a factor of $\sim 2.5$ for every magnitude
deeper \cite{Tyson88}.  A medium deep observation capable of shape
measurements to $R \sim 25$ thus yields about 20 galaxies
arcmin$^{-2}$ (assuming a bright cutoff $R>23.5$ to eliminate largely
foreground sources), implying that 2 arcmin$^{2}$ are required for 5\%
accuracy in shear.  \index{source!density}

Getting more sources per unit area requires much more telescope time.
Source density will ultimately be limited by confusion 
\index{source!confusion} --- when sources
are so numerous that they overlap and hinder shape
measurements --- around $\sim 1000$ sources arcmin$^{-2}$ for
ground-based data.  This implies $\sim$ 20\arcs shear resolution, or
better for space-based data, if galaxy counts keep rising at the same
rate.  However, such depth is hard to come by and must compete against
area and wavelength coverage (useful for constraining source
redshifts) when planning for a given amount of telescope time.

Another tradeoff commonly used is to sacrifice resolution in one
dimension to achieve better resolution in the other.  Clusters are
commonly analyzed in terms of a radial profile, which assumes they are
axisymmetric and allows all sources at a given radius from the cluster
center to be averaged together.  Less massive clusters and groups can
be ``stacked'' to yield an average profile with reasonable resolution,
just as in galaxy-galaxy lensing \cite{Hoekstra_etal2001,Sheldon}.  

\subsubsection{Source redshift distribution}
\label{sec-srd}
\index{source!redshift distribution}

Lack of knowledge of the source redshift distribution is often a limit
in calibrating weak lensing measurements.  The root of this problem is
that deep imaging quickly outruns the ability of even the largest
telescopes to provide spectroscopic redshifts for a fair sample of
sources.

The recent development of photometric redshift techniques, 
\index{photometric redshift} in which
multicolor imaging provides enough spectral information for a
reasonable redshift estimate cite{Connolly1995,Hogg1998}, has brought
hope that source redshifts may be estimated to sufficient accuracy
from imaging alone.  For example, the Hubble Deep Field \index{HDF} yielded
photometric redshifts accurate to $\sim 0.1$ per galaxy in the redshift range
$0-1.4$ with seven filters extending through the near-infrared
($UBVIJHK$) \cite{Hogg1998}.  A look at Figure~\ref{fig-dratio} shows that this
provides a reasonable accuracy in distance ratio in most situations.
The accuracy improves with the number of filters used, resulting in a
tradeoff between accuracy and telescope time.  Deep $U$ and infrared
imaging are much more expensive than $BVRI$ in terms of telescope
time, but it is difficult to effectively cover a large redshift range
with only $BVRI$.  Few spectral features are to be found in the
observed $BVRI$ bandpasses for sources in the redshift range $\sim
1.5-3$, which greatly increases uncertainties there.

However, these problems are not fundamental, and photometric redshifts
will become routine.  They will do much more than help estimate the
mean distance ratio required for calibrating lenses.  Because sources
lie at a range of redshifts, they will provide the opportunity to
probe structure along the line of sight (albeit with resolution
limited by the width of the lensing kernel).  The ultimate goal is
{\it tomography} \index{tomography} --- 
building up a three-dimensional view of mass in the
universe from a series of two-dimensional views at different
redshifts.  The combination of weak lensing and photometric redshifts
thus promises to be very powerful, but as yet there are not many
published examples of combining the two, and little theoretical work
on optimal ways of doing so.  Although we can expect photometric
redshifts \index{photometric redshift} 
to be a routine part of future lensing work, we must be
aware of alternative ways of confronting the source redshift problem.
\index{source!redshift}


First, some questions can be answered without calibration of source
redshifts.  The two-dimensional morphology of a cluster lens is one
example  --- the source redshift distribution should not depend on
position (as long as magnification is negligible and cluster members
do not contaminate the source sample).  Similarly, source redshifts
are not required for discovery of mass concentrations in surveys, but
without them, the volume probed is unknown.  Clearly, the questions
which can be answered this way are limited.

A more general calibration strategy is through additional, identical
observations of a ``control lens'' of known redshift and mass ({\it
e.g.} a cluster with a dynamical, X-ray, \index{cluster!X-ray} 
and/or strong lensing mass
estimate).  This does allow estimation of the mean distance ratio to a
population of sources much too faint to reach with spectroscopy, but
it certainly has its limits.  It is difficult to obtain identical
observations, and the (probably considerable) uncertainty in the mass
of the control lens becomes a systematic for the rest of the data.
But more fundamentally, shear from the control lens samples only that
part of the source distribution which is behind the control lens, so
that strictly speaking, a control lens must be at the same redshift as
the target.  For weak lensing by large-scale structure, \index{large-scale
structure} the
distribution, not simply the mean distance ratio, is required.  This
would require control lenses at a range of redshifts, which is
impractical.  Photometric redshifts should do a much better job with
more realistic data requirements.  Even in the age of photometric
redshifts, though, this method will have its role.  The shear induced
by calibrated lenses will provide a check on photometric redshift
estimates, which may not be checkable with spectroscopy if applied to
very faint sources.

Another strategy is keeping the imaging as shallow as current redshift
surveys, which go to $R \sim 24$.  One can then look up the median
redshift for any magnitude cut; for $23<R<24$, for example, the median
redshift is 0.8 \cite{redshiftsurvey}.  Even the redshift distribution
is known to some extent, with 120 sources in that magnitude slice in
the survey cited.  Shallow imaging need not probe a small volume, as a
large area can be covered with a reasonable amount of telescope time.
But it does limit the distance probed and the angular resolution of
the mass reconstruction
(because the areal density of sources is low at $R \le 24$).  This
strategy also limits selection of sources based on color, which is
very useful for limiting contamination by galaxies in a cluster being
studied (or for de-emphasizing foreground contamination in general)
because the median redshift of a color-selected sample is not yet
something that can be looked up in a redshift survey.

%
%

\subsubsection{Intrinsic alignments}
\index{source!alignment} 

The crucial assumption in weak lensing is that the sources have random
intrinsic orientations, so that any departure from randomness is due
to lensing.  This assumption is worth examining before proceeding
further.  We will concentrate on potential damage to measurements of
weak lensing by large-scale structure (cosmic shear), because the
lensing signal from clusters is usually at a much higher level.
However, it is worth keeping in mind that all applications of weak
lensing could be affected at some level.

The first detections of cosmic shear \index{shear!cosmic} 
in 2000 motivated several
analytic \cite{CKD2000} and computational \cite{HRH2000,CM2000} studies
of intrinsic alignment \index{source!alignment} 
mechanisms, and the field is still sorting
itself out.  There are several mechanisms which could produce such
intrinsic alignments, including tidal stretching of galaxies in a
gravitational potential, and coupling of the potential to the spin
vectors of galaxies \cite{CNPT2000a}.  The amount of alignment
predicted as a function of angular scale varies greatly depending on
the mechanism and the strength of the coupling; it remains unknown
which model, if any, is correct.  However, in most scenarios,
intrinsic alignments would represent a
\raisebox{-0.5ex}{$\stackrel{<}{\scriptstyle \sim}$} 10\%
contamination of the cosmic shear measurements.

\begin{figure}
\centerline{\resizebox{2.2in}{!}{\includegraphics{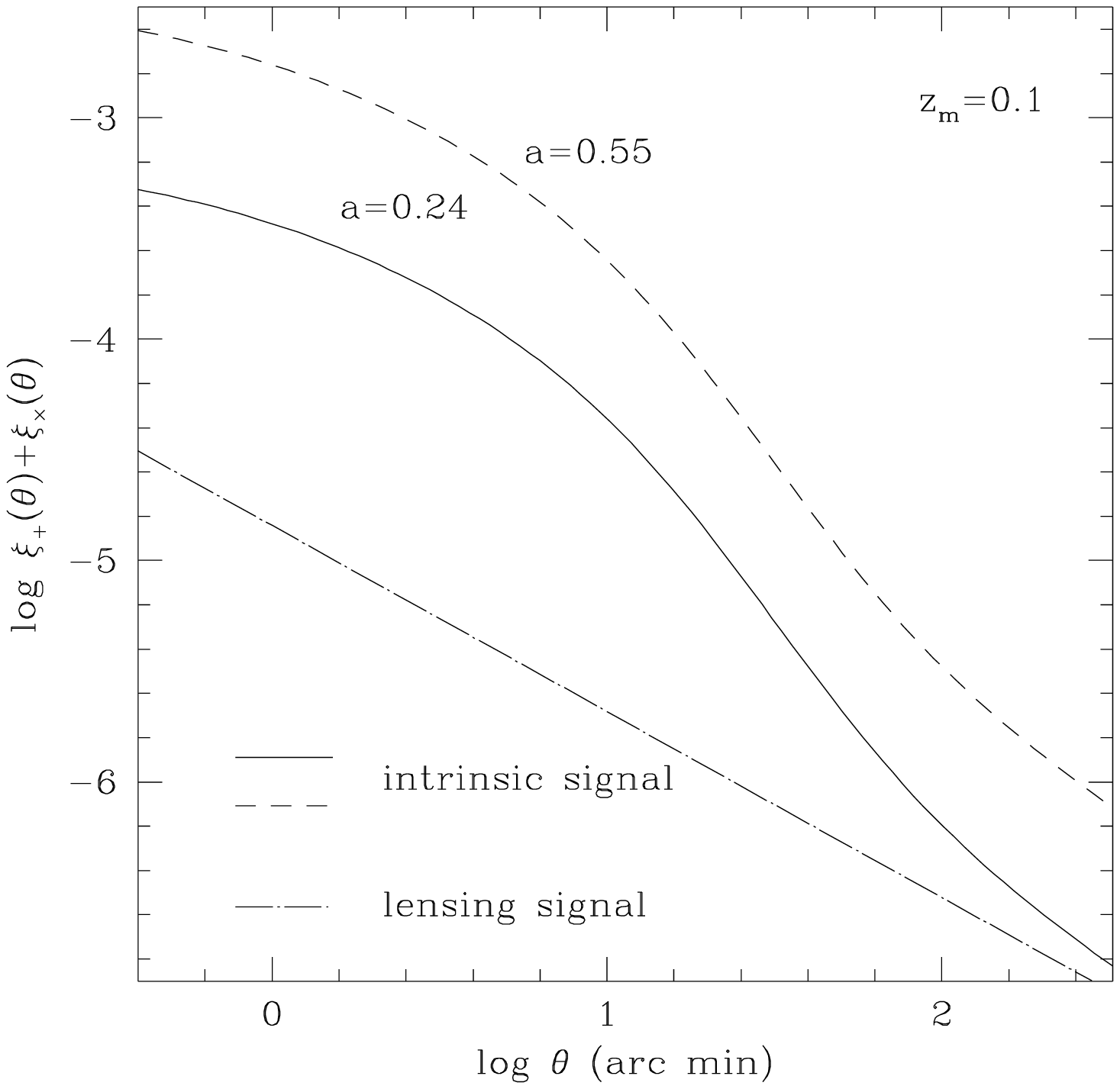}}
\resizebox{2.2in}{!}{\includegraphics{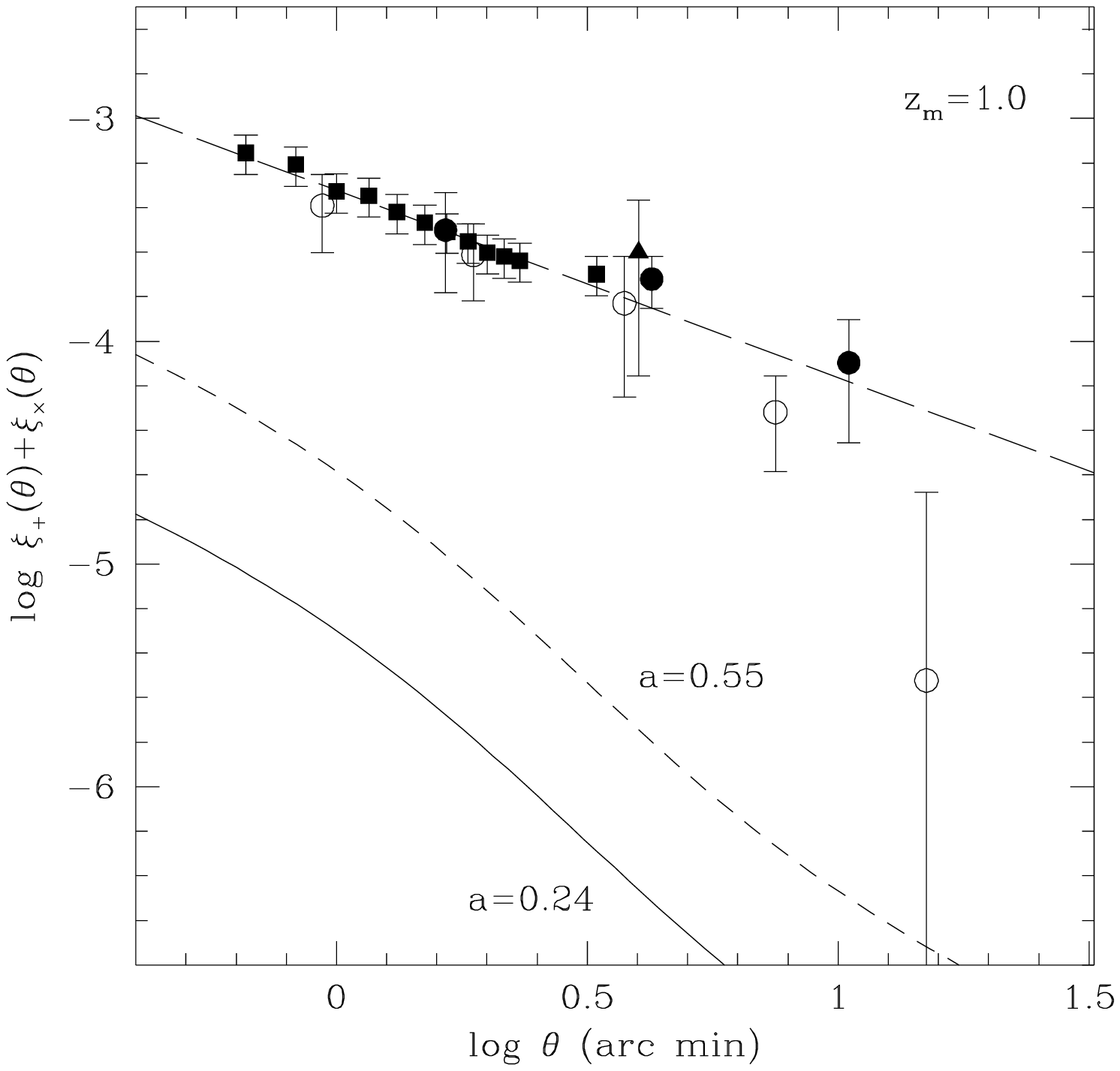}}}
\caption{ The importance of intrinsic alignments 
depends strongly on \index{source!redshift} \index{source!alignment}
source redshift.  The expected levels of ellipticity correlation
(defined in Section~\ref{sec-cosmicshearstats}) due to intrinsic
alignments and to weak lensing by \index{large-scale structure} 
large-scale structure are shown for
low-redshift (median source redshift $z_m = 0.1$) surveys in the left
panel and for high-redshift ($z_m = 1$) surveys in the right panel.
In each case, the straight line indicates the expected signal from
weak lensing, and the curves indicate the expected signal from
intrinsic alignments, for two different values of a spin-coupling
parameter, giving some idea of the modeling uncertainties.  The right
panel also contains cosmic shear \index{shear!cosmic} 
measurements from the literature,
which all happen to have $z_m \sim 1$.  Adapted from
 \cite{CNPT2000a}.}
\label{fig-cnpt1}
\end{figure}

While the situation is still evolving, one rule is certain: the effect
of any intrinsic alignment is diluted when sources lie at a large
range of redshifts, as is naturally the case in deep imaging.  As we
shall see, the signal from lensing by large-scale structure 
\index{large-scale structure} increases
with source redshift.  Hence, lensing must dominate at high
enough source redshift, and intrinsic alignments at low enough
source redshift.  This is illustrated by Figure~\ref{fig-cnpt1}, which
shows predicted intrinsic alignment and cosmic shear levels for
several source redshifts.  For shallow surveys like the Sloan Digital
Sky Survey \cite{SDSS} (SDSS), \index{SDSS} intrinsic alignment may frustrate
attempts to measure cosmic shear, but deeper surveys specifically
designed to measure cosmic shear are safe.  Indeed, the roughly one
million spectroscopic redshifts SDSS plans to acquire will be
invaluable in measuring intrinsic alignments precisely, and their
measurements, after scaling to higher redshifts, in turn may
facilitate estimation and even removal of intrinsic alignment effects
from the deeper surveys.  Deep surveys may also be able to provide a
lensing signal using only sources which cannot be physically
associated, as indicated by their photometric redshifts.  Density
reconstruction methods in the presence of intrinsic alignments are
already being investigated \cite{LP2000}.

The two or three detections of intrinsic alignments in real data are
indeed at low redshift.  Ellipticity correlations have been reported
in SuperCOSMOS data \cite{BTHD2000}, but because there is no redshift
information, intrinsic alignments can only be inferred (the median
source redshift is estimated to be $<0.1$, so that the inferred cause
is intrinsic alignments rather than lensing).  Spin alignments have
been found in the Tully catalog, which consists of several thousand
nearby (within a few Mpc) spirals \cite{PLS}, and in the PSCz, a
redshift survey of 15500 galaxies detected by the IRAS infrared \index{IRAS}
satellite mission \cite{LP2001}.  Because these catalogs have
redshift information, these represent solid detections.  However, spin
correlations are one step removed from ellipticity correlations, which
are the relevant quantity for lensing.

It may also be possible to extract intrinsic alignments from the
lensing data itself.  To first order, lensing produces a curl-free, or
E-type (in analogy with electromagnetism) shear field.  (Multiple
scattering can produce a weak divergence-free, or B-type field, but
that can be safely ignored for the moment.)  Therefore, decomposition
of a measured shear field into E-type and B-type fields might allow
separation of the lensing and intrinsic alignment
effects \cite{CNPT2000b}.  This decomposition is difficult, but a crude
indicator of the B-type field is the traditional 45$^\circ$ test.  In
this test, a lensing signal should vanish when one component of the
shear is exchanged for the other (equivalent to rotating each source
by 45$^\circ$), and nonzero results would indicate a systematic error.
All published cosmic shear results were vetted using this
test among others, with no indication of contamination.  However, not
all intrinsic alignment mechanisms produce B-type power; an example is
tidal stretching (tidal fields are the basic mechanism for both
stretching and lensing, after all).  Based on other astrophysical
arguments, tidal stretching is not likely to be
significant \cite{CNPT2000b}, but even angular momentum coupling models
can produce much more E-type than B-type correlations \cite{MWK2001}.
Passing the 45$^\circ$ test is a
necessary but not sufficient condition for confidence in the results.

In summary, intrinsic alignments are not to be dismissed.  They must
be addressed and may even dominate the lensing signal in certain
low-redshift applications.  However, the dilution effect of a broad
source redshift distribution \index{source!redshift distribution} 
means that none of the conclusions of
weak lensing to this point can be called into doubt.  Ongoing and
future weak lensing studies may have to apply small corrections for
this effect, but how small is still uncertain.  Accurate corrections
will probably be available by the completion of the SDSS, \index{SDSS} 
which will do
much to increase our knowledge of intrinsic alignments in the nearby
universe.


\subsection{Measuring shear}
\index{shear!measurment}


In most weak lensing work, a source galaxy is approximated as an
ellipse fully described by its quadrupole moments 
\index{source!ellipticity}
\begin{eqnarray}
I_{xx} \equiv { \Sigma Iwx^2 \over \Sigma Iw } \nonumber \\
I_{yy} \equiv { \Sigma Iwy^2 \over \Sigma Iw } \\
I_{xy} \equiv { \Sigma Iwxy \over \Sigma Iw } \nonumber 
\end{eqnarray}
where $I(x,y)$ is the intensity distribution above the night sky
level, $w(x,y)$ is a weight function, the sum is over a contiguous set
of pixels defined as belonging to the galaxy, and the coordinate
system has been translated so that the first moments vanish ({\it
i.e.} the centroid of the galaxy is chosen to be the origin of the
coordinate system).  Early work used intensity-weighted
moments ($w=1$), but it was realized that this produces ellipticity
measurements with noise properties that are far from optimal or even
divergent.  Now, $w$ is usually chosen to be a circular \cite{KSB} or
elliptical \cite{BJ2001} Gaussian, which deweights the outer pixels
which have a big lever arm but low signal-to-noise.  The two
ellipticity components \index{ellipticity!components} 
can be defined as \cite{VJT83}
\begin{eqnarray} 
e_+ = { I_{xx} - I_{yy} \over I_{xx} + I_{yy}} \nonumber \\
e_\times = { 2I_{xy} \over I_{xx} + I_{yy}} 
\end{eqnarray}
These are related to the scalar ellipticity $\epsilon$ and position
angle $\phi$ by \index{source!position angle}
\begin{eqnarray} 
\epsilon = (e_+^2 + e_\times^2)^{1 \over 2} \nonumber \\
\phi = {1 \over 2} tan^{-1}({e_\times \over e_+}) 
\end{eqnarray}
Then a simple estimate of the shear in the weak lensing limit 
\index{weak lensing!limit} is
$\gamma_i = \langle e_i \rangle/2$, where the brackets denote
averaging over many sources (perhaps with weighting of the sources
based on estimated measurement errors, redshift, etc.) to beat down
shape noise. \index{shape noise} 
Note that this definition of ellipticity differs from
that in Equation (\ref{eq-shear}) by a factor of two; both definitions
are presented here because both are common in the literature.  This
latter estimator sometimes called the {\it distortion} statistic.
Also, there are alternative formulations in terms of octupole
moments \cite{octupole}, Laguerre expansions \cite{BJ2001} and
shapelets \cite{shapelets1,shapelets2,shapelets3}.
Before applying any of these estimators, we must account for 
the effects of point-spread function (PSF) anisotropy \index{PSF!anisotropy} 
and broadening.


\subsubsection{PSF anisotropy}
\index{PSF!anisotropy}
\index{PSF!ellipticity}

No optical system is perfect, and PSFs on real telescopes tend to be
$\sim 1-10\%$ elliptical.  This constitutes a huge systematic error,
of the order of the shear induced by even a massive cluster, and it
must be removed as completely as possible before analyzing any galaxy
shapes, and monitored afterward.  The removal can be done {\it after}
measuring shapes, by essentially subtracting the moments of the PSF
from the galaxy moments, but a more computationally stable method is
to remove this effect from the image, {\it before} measuring shapes.
This is done by convolving the image with a kernel with ellipticity
components \index{ellipticity!components} 
opposite to that of the PSF \cite{FT97}.  The raw PSF is
almost certainly position-dependent; therefore the circularizing
kernel is also, but the convolved PSF is everywhere round.  A round
PSF is called {\it isotropic}, but keep in mind that this does not
imply homogeneous: The convolved PSF may vary somewhat in size,
because of the position dependence of the original PSF and of the
small broadening introduced by the kernel.  A more sophisticated
scheme would introduce more broadening in the right places, leading to
a PSF which is homogeneous as well as isotropic.
It is also possible to
choose a sharpening kernel, but this would amplify the noise in the
image.

\begin{figure}[ht]
\centerline{\resizebox{3in}{!}{\includegraphics{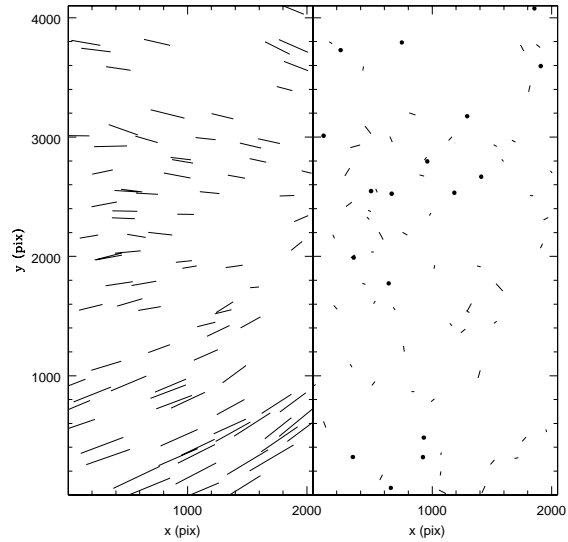}}}
\caption{Point-spread function correction in one 2k$\times$4k CCD.
Shapes of stars, which as point sources should be perfectly round, are
represented as sticks encoding ellipticity and position angle.  Left
panel: raw data with spatially varying PSF ellipticities up to 10\%.
Right panel: after convolution \index{PSF!convolution} 
with a spatially varying asymmetric
kernel, ellipticities are vastly reduced (stars with $\epsilon<0.5\%$
are shown as dots), and the residuals are not spatially correlated as a
lensing signal would be.}
\label{fig-circ}
\end{figure}

\begin{figure}[ht]
\centerline{\resizebox{3in}{!}{\includegraphics{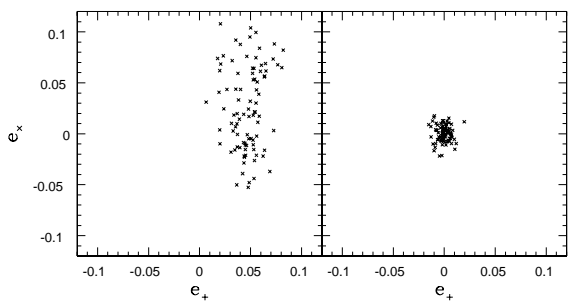}}}
\caption{Another way of plotting the efficacy of the PSF correction,
\index{PSF!correction} 
often seen in the literature.  For the same dataset shown in
Figure~\ref{fig-circ}, the ellipticity components of point sources are
shown in a scatterplot, before and after correction.  This type of
plot hides any spatial correlation which may exist among the
residuals, but Figure~\ref{fig-circ} shows that the residuals are
uncorrelated in this case.}
\label{fig-escatter}
\end{figure}

Figures~\ref{fig-circ} and~\ref{fig-escatter} illustrate the
effectiveness of the convolution procedure.  Although there are
low-level residuals in the convolved image, their lack of spatial
correlation means that they will have difficulty masquerading as a
weak lensing effect.  Note that these PSF anisotropies change with
time, as telescope temperature, focus, and guiding drift, so that each
exposure must be treated separately.  A possible benefit here is that
if the anisotropies are really uncorrelated temporally, coaddition of
multiple exposures will beat down the shape errors.  Also, each CCD in
a mosaic must be treated separately, as some discontinuities may arise
from small differences in piston between devices.

\subsubsection{PSF broadening}
\index{PSF!broadening}

Any effect which broadens the PSF will reduce the measured
ellipticities of source galaxies which are not much larger than the
PSF---generally including the distant galaxies most appropriate for
lensing---because they will be broadened relatively more along their
minor axes than along their major axes.  In ground-based data, the
dominant effect is ``seeing'', the broadening of the point-spread
function due to turbulence in the atmosphere.  Note that this is
distinct from PSF anisotropy, which is caused by the telescope and
camera optics.  In fact, seeing produces a circular PSF as long as the
integration time is much longer than the coherence time of the
atmosphere (very roughly 30 ms at visible wavelengths); anyone who has
observed in terrible seeing has probably noticed that at least the PSF
is nicely round !  The effects of PSF anisotropy and broadening are
sometimes called ``shearing'' and ``smearing'', respectively.  The
former has the effect of introducing a spurious weak lensing signal if
uncorrected, while the latter has the effect of reducing any weak
lensing signal.

There are several ways of correcting for smearing.  The first is
measuring the dilution of a simulated weak lensing signal relative to
an unsmeared image, either simulated or perhaps from the Hubble Deep
Fields. \index{HDF} The seeing-free images are sheared by a known amount,
convolved with the point-spread function of the real data,
repixelized, and the shear measured.  This has the advantage of
including some effects which cannot be accounted for analytically,
such as the coalescing of separate objects into an apparently
elliptical single object.

On the other hand, a global correction is rejected by those who prefer
to tailor the corrections to individual sources; after all, a large
galaxy is smeared relatively less by seeing than is a small one.  The
advocates of this approach tend to use the KSB method, an analytical
approach which takes into account the size of the PSF and of each
source \cite{KSB}.  The KSB method also accounts for PSF shearing, but
it can just as well be applied to a convolved image.  See
 \cite{Kaiser1999,Kaiser2000,Kuijken,BJ2001} for limitations of and
possible successors to the KSB method.  These approaches also weight
each source according to its ellipticity uncertainty when computing a
shear estimate.

\subsubsection{Source selection}

Not every source in a deep image should be included in a shear
measurement.  A typical deep image includes stars, other unresolved
sources, foreground galaxies, cluster members if the target lens is a
cluster, and spurious objects, such as bits of scattered light around
very bright stars.  Getting rid of these unwanted sources is something
of an art, which must reflect the particular data set, but generally
there are four kinds of cuts.  Magnitude cuts help get rid of stars
(for reasonable galactic latitude, stars outnumber galaxies for $R$
\raisebox{-0.5ex}{$\stackrel{<}{\scriptstyle \sim}$}~$22$ while
galaxies greatly outnumber stars for $R > 23$) and bright foreground
galaxies.  Galaxies have a broad luminosity function, so such a cut is
never completely effective at eliminating the foreground, but it
helps.  Color cuts seek to emphasize the faint blue galaxies at $z
\sim 1$  \cite{TysonSeitzer1988}.  If the target is a cluster, the cut
should be blueward of the cluster's color-magnitude ridge.  Even so,
some cluster members and other foreground galaxies will survive.  Size
cuts eliminate unresolved objects, which at the relevant magnitudes
include some stars, but mostly unresolved galaxies.  Finally, cuts
designed to insure that an object is not spurious must depend on the
type of data available.  Examples include rejecting objects that
appear on only one of a multicolor set of images, and rejecting high
ellipticity objects which are likely to be unsplit superpositions of
two different objects.

\subsubsection{Sanity checks}
There are a number of sanity checks that should be performed before
believing any weak lensing result.  In addition to the 45$^\circ$ test
mentioned above, randomizing source positions while retaining their
shapes should result in zero signal.  Another good sanity check is
correlating the source shapes with an unlensed control population,
such as a set of stars.  Finally, there are checks on the basic
integrity of the catalog, such as the position angle distribution of
sources, which might reveal spurious objects aligned with the detector
axes.  Because setting the source selection criteria can be somewhat
subjective, it is also good to check that the results do not depend
crucially on the exact magnitude or color cut.

\section{Lensing by clusters and groups}
\index{galaxy!cluster} 
\index{galaxy!groups}
\index{cluster!as a lens}

Clusters of galaxies have long been studied from two somewhat opposing
points of view.  Visible from great distances, they are a convenient
tracer of structure in the universe back to roughly half its present
age.  When examined individually, they are interesting astrophysical
laboratories in their own right, with a variety of physical conditions
and histories.  But if so, they cannot be simple cosmological probes.
So the study of clusters as astrophysical laboratories must inform and
refine the study of clusters as cosmological probes.
\index{cluster!dynamics}
\index{cluster!X-ray}

What lensing adds to the study of clusters is a direct mass
measurement without any assumptions about the dynamical state of the
cluster.  The first clusters were ``weighed'' in the 1930's
with the dynamical method---assuming that clusters are in virial
equilibrium, the virialized mass is easily computed from the velocity
dispersion.  In the late 1980's, X-ray imaging of hot intracluster gas
began to provide mass estimates, assuming hydrostatic equilibrium.  In
the 1990's, lensing began to provide mass estimates free of any such
assumptions.  The frequent agreement of the three types of estimate
indicates that the dynamical assumptions are often valid, but the
exceptions need to be identified.  Those exceptions must be discarded
from any samples used as cosmological probes, but they are often
studied more closely for what they might reveal about mergers or other
nonequilibrium processes.

In the past decade, it was enough simply to compare lensing
measurements of cluster masses with those provided by other techniques.
Driven by advances in wide-field detectors, we can now use lensing to
{\it search} for clusters (or at least mass concentrations), and even
estimate their redshifts.  Shear-selected samples of clusters, free of
any bias toward baryons that optically and X-ray selected samples
might have, are currently being compiled.  Comparison of the different
types of samples will be instructive, either by confirming the use of
traditional baryon-selected samples as cosmic probes, or perhaps by
providing some counterexamples.

%

\subsection{Masses and profiles}
\index{mass!profile}

\begin{figure}
\centerline{\resizebox{3in}{!}{\rotatebox{-90}{\includegraphics{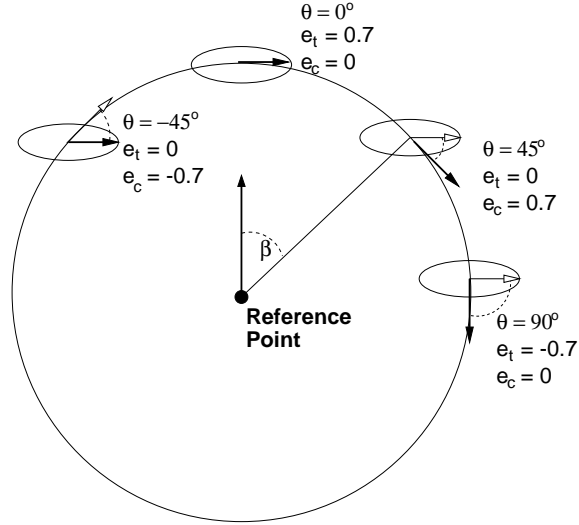}}}}
\caption{Tangential ellipticity $e_t$ of an $\epsilon = 0.7$ source
with respect to a reference point.  $e_t$ carries the lensing signal,
and the 45-degree component $e_c$ serves as a control.  In the
presence of lensing but not shape noise, all sources would have $e_t >
0$ and $e_c = 0$; in practice $\langle e_t \rangle > 0$ and $\langle
e_c \rangle = 0$. \index{ellipticity!components}}
\label{fig-tangential}
\end{figure}

The first evidence of lensing by clusters came in the late 1980's in
the form of strongly lensed giant arcs \cite{LP86,Soucail87}, 
\index{arcs} \index{arclets} which
were used to constrain the mass inside the radius at which the arcs
appeared.  This was soon extended to somewhat less strongly lensed
``arclets'', and by 1990, to the first detection of what we now call
weak lensing, the coherent alignment of thousands of weakly lensed
background galaxies \cite{TWV90}.  This alignment was measured in terms
of the {\it tangential shear} $\gamma_t$, which is the component of
\index{ellipticity!components} 
shear directed tangential to an imaginary circle centered on the
cluster and running through the source.  The tangential
ellipticity of a source is $e_t \equiv \epsilon
\cos (2\theta)$ where $\theta$ is the angle from the tangent to the
major axis of the source (Figure~\ref{fig-tangential}).
When computing $e_t$ from the ellipticity components, use the
rotation
\begin{eqnarray}
e_t = +e_+ \cos(2 \beta)  +  e_\times \sin(2 \beta) \nonumber \\
e_c = -e_+ \sin(2 \beta)  +  e_\times \cos(2 \beta)
\end{eqnarray}
where the angle $\beta$ is also shown in Figure~\ref{fig-tangential}.
Here $e_c$ is a control statistic measuring the alignment along an
axis 45 degrees from the tangent, which is not affected by an
axisymmetric lens.

Methods for constraining cluster masses using tangential
shear
followed soon after the first detection of the
effect \cite{Jordi1991a}.  The most important of these is {\it aperture
densitometry}, which relates $\gamma_t$ to the difference between the
mean projected mass density inside a radius $r_1$ and that between
$r_1$ and a larger radius $r_2$ \cite{Fahlman1994}:
\begin{equation}
 \bar{\kappa}(<r) - \bar{\kappa}(r_1<r<r_2) = {2 \over 1 -
r_1^2/r_2^2} \int^{r_2}_{r_1} {\gamma_t \over 1 - \kappa(r)} d \ln r.
\end{equation}
The factor $1-\kappa$ can be ignored in the weak lensing limit, 
\index{weak lensing!limit} but \index{cluster!as a lens}
for massive clusters may not be ignored, leading to an iterative
solution for $\kappa$ ({\it e.g.}  \cite{FT97}).  The left hand side of
this equation is sometimes called the {\it zeta statistic}
$\zeta(r_1,r_2)$. Note that this formula makes the mass sheet
degeneracy \index{degeneracy!mass sheet} 
explicit by specifying only relative values of $\kappa$;
the best that can be done is extend $r_2$ to a very large value, at
which $\kappa$ should vanish.

A profile can be built up by repeatedly applying this statistic at a
sequence of different $r_1$.  However, this makes the points in the
profile dependent on each other, as they use much the same data.  If
the goal is to find the best fit of a given type of profile, it is
simpler to compute $\gamma_t$ in a series of independent annuli and
fit the shear profile expected from the mass model straightforwardly
with least-squares fitting. 

\begin{figure}
\centerline{\resizebox{3in}{!}{\includegraphics{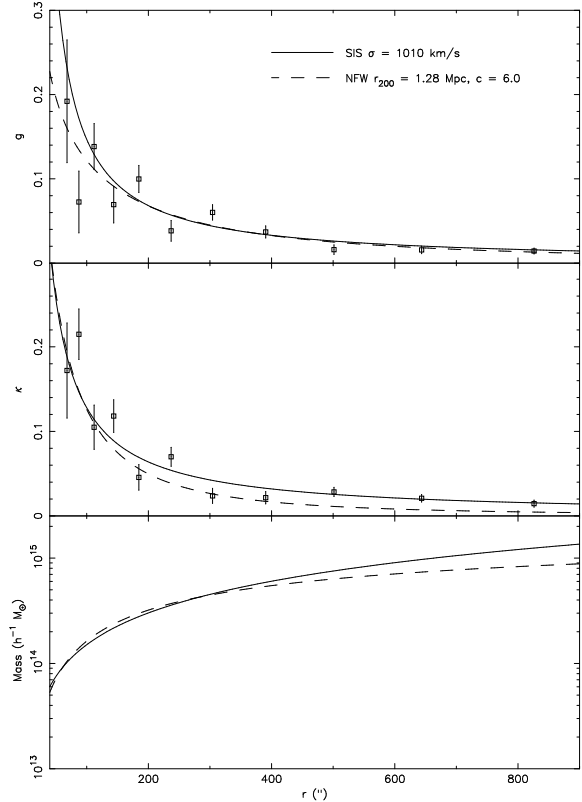}}}
\caption{ Comparison of shear, convergence, and mass profiles for the
massive cluster Abell 1689.  {\it Top:} (Reduced) shear profile, with
best-fit SIS and NFW profiles almost indistinguishable.  {\it Middle:}
$\kappa$ profiles, showing that the NFW model falls off much more
steeply than the SIS at large radii.  This could barely
be seen in the shear profile because of the nonlocality of shear.
Despite appearances, the data do not favor the SIS.  The points here
are plotted assuming that outside the largest radius measured, the
shear falls off as an SIS.  If NFW is instead assumed for radii
outside the measurement area, the points at large radius fall
significantly.  {\it Bottom:} Enclosed mass profiles of the two
models. The physical scale is roughly 2 kpc arcsec$^{-1}$.  From
 \cite{CS2001}. \index{lens!Singular Isothermal Sphere}}
\label{fig-profiles}
\end{figure}

Weak lensing mass profiles are usually well fit by a singular
isothermal sphere (SIS) \index{lens!Singular Isothermal Sphere} 
or Navarro-Frenk-White (NFW,
 \cite{NFW95,NFW97}) profile (see  \cite{Keeton2001} for an extensive
list of profiles used in lensing, along with their associated
formulae).
However, the nature of weak lensing makes it difficult to distinguish
between models on two accounts.  First, shear profiles do not have
good dynamic range because the uncertainty in shear measurements
increases dramatically at small radii, where there are not enough
sources to beat down the shape noise. \index{shape noise} 
This is illustrated in the top
panel of Figure~\ref{fig-profiles}.  Note that this figure is for a
very massive cluster; the signal-to-noise ratio can only be lower for
less-massive clusters.  Second, mass profiles which differ
significantly {\it inside} the radius where shear is measured can
produce shear profiles which differ significantly {\it only outside}
that radius.  This is illustrated in the middle panel of
Figure~\ref{fig-profiles}.  The ability to distinguish between NFW and
SIS (or more generally, power-law profiles) thus depends strongly on
\index{cluster!as a lens}
the size of the field \cite{KS2000}, but Figure~\ref{fig-profiles}
demonstrates that a significant ambiguity remains even with a
state-of-the-art imager with a 35' field.  Weak lensing is therefore
not definitively revealing cluster mass profiles, as one
might have expected.  Progress toward larger fields will be slow, as
most large telescopes already have imagers which fill their usable
fields of view.  More likely, progress will come by adding
magnification information.  Finally, note that the most active (and
revealing of the nature of dark matter) controversy surrounding
cluster profiles involves cuspiness at the center, and this is not
well addressed by weak lensing, with its poor angular resolution.


Initially, weak lensing analyses of clusters concentrated exclusively
on the most massive clusters which were guaranteed to give a good
signal.  
As the technique has matured, it has been extended
to less massive, but more typical clusters \cite{NCS}.
The ultimate extension has been to groups; although a group by itself
does not provide enough shear to get an accurate mass estimate, they
can be ``stacked'' as in galaxy-galaxy lensing to build up an average
profile with reasonable signal-to-noise \cite{Hoekstra_etal2001}. The
idea of stacking to obtain a good estimate of the average profile has
been used for typical clusters as well \cite{Sheldon}. Caution is
required when interpreting ``average'' results, though, because they may be
biased by a few unrepresentative systems, or in the worst case,
meaningless if the sample is sufficiently heterogeneous.

Mass estimates of clusters and groups 
\index{galaxy!cluster} \index{galaxy!groups} derived from weak lensing
generally agree with estimates from velocity dispersions and X-ray
imaging (see  \cite{Mellier} for a list of published mass estimates as
of 1999; there are now too many to list).  At one time there was an
apparent systematic discrepancy between (strong) lensing and X-ray
\index{cluster!X-ray}
estimates \cite{MB95}, but it was shown to be due to the complexities
of the X-ray-emitting gas dynamics \cite{Allen1998}. Hydrostatic
equilibrium alone was shown to be less constraining than initially
thought; temperature maps were needed \cite{BB97}.  XMM and Chandra now
provide these, along with vastly improved angular resolution which
allows for better treatment of cooling flows, and the first results
show good agreement with lensing \cite{ASF2001}.  Of course, not every
cluster behaves so well, and when there is disagreement a closer look
often reveals interesting astrophysics such as cooling flows, mergers
and their associated ``cold fronts'' or shock heating.  In fact,
$\sim$50\% of clusters show some substructure in the
X-ray \cite{Schuecker}.  Weak lensing can still supply the total mass,
but due to its poor angular and line-of-sight resolution, the detailed
work of disentangling the structures must be left to X-ray and
dynamical measurements.

The three approaches, after all, have some fundamental differences
which are not often mentioned. Dynamical estimates based on the virial
theorem measure the total virialized mass of the cluster, while
lensing can only measure mass projected inside a certain radius.  Even
with a simplifying assumption such as spherical symmetry, a fair
comparison of the two is difficult.  Virial masses go as the square of
the velocity dispersion, so small-number statistics and outliers can
have a large effect on the mass \cite{Saslaw}.  Mass estimates from a
dozen members may be good for a back-of-the envelope comparison, but
beyond that should be treated with extreme caution.  Velocity {\it
profiles} would be more comparable to lensing and X-ray data.  These
are available for few clusters, but thanks to multi-object
spectrographs on large telescopes, such detailed dynamical analyses
are becoming more common \cite{Biviano2001}.  X-ray emission is
proportional to the square of the density, so it is more
sensitive to substructure than is lensing, which is simply
proportional to the density.  A fourth approach, the
Sunyaev-Zel'dovich \index{Sunyaev-Zel'dovich} 
effect (SZE), measures the decrement in the cosmic
microwave background (CMB) \index{CMB} 
caused by upscattering of CMB photons by \index{cluster!X-ray} 
the hot intracluster gas.  Like lensing, it is proportional to the
density, but like X-ray emission, it depends on the density of
baryons, rather than all matter.  The first SZE measurements are
starting to arrive and will soon offer their unique point of view.

Lensing stands out from X-ray and dynamical methods in being a
projected statistic, so it is worth asking whether this introduces any
bias.  It appears that anisotropy in simulated clusters has little
systematic effect \cite{Brainerd99}, and so do uncorrelated structures
along the line of sight \cite{Hoekstra}; both effects are around the
5\% level.  However, in reality there are also correlated structures
along the line of sight, and these can bias masses upwards by tens of
percent  \cite{Cen97,MWL2001}. There is general agreement that the
effect of other structures along the line of sight increases with
aperture size.  The bias changes with redshift in two ways.  First, it
is minimized when the cluster is near the peak of the lensing kernel
(Figure~\ref{fig-efficiency}), because other structures will be
deemphasized.
Second, younger clusters may have more nearby material, although this
effect has not been investigated thoroughly \cite{MWL2001}. Finally,
note that the mass function is susceptible to bias even when an
estimator is unbiased but has scatter, because there are more low-mass
clusters to be scattered up than high-mass clusters to be scattered
down (this applies equally to other types of mass estimates such as
dynamical and X-ray) \cite{MWL2001}.

\subsection{Two-dimensional structure}
\index{convergence}
\index{shear}

From the first detection of weak lensing, it was realized that the
tangential shear procedure could be repeated about any reference
point, not only the cluster center.  By repeating it at a grid of
points, a two-dimensional ``mass map'' \index{mass!map} 
(actually a map of $\kappa$)
was constructed \cite{TWV90}.  This was soon put on a firm theoretical
footing by the derivation of a relationship between the Fourier
transforms of $\kappa$ and $\gamma$, starting from their relationship
as different linear combinations of the second derivatives of the
lensing potential $\psi$ \cite{KS93} (see the Quasar Lensing chapter
for the relationship between $\psi$ and $\kappa$).  Essentially,
$\kappa$ can be expressed as a convolution of $\gamma$ over the entire
plane (there is also a real-space equivalent \cite{FT97}).  Of course,
observations do not cover the entire plane, a problem called the {\it
finite field effect}  \cite{Bartelmann95}.  This is another
manifestation of mass sheet degeneracy, \index{degeneracy!mass sheet} 
as the spatial variation, but
not the mean value, of $\kappa$ can be reconstructed.  Several
reconstruction methods based on magnifications have been proposed to
combat this problem \cite{BTP95,BN95}, but, as mentioned above,
magnification is very difficult to measure, and these methods have not
been widely used.  To a large extent, technology has solved the
problem, at least for clusters, by providing ever-wider fields of
view, at the edges of which $\kappa$ is presumably negligible.

In addition to direct {\it reconstruction} methods, there are {\it
inversion} methods which solve for $\psi$, from which $\kappa$ can be
derived \cite{Bartelmann1996}. An extensive comparison of different
methods found none to be clearly superior \cite{SK96}, although
inversion methods tend to make it easier to include additional
constraints such as those from magnification measurements or strong
lensing features.

\begin{figure}
\centerline{\resizebox{3in}{!}{\includegraphics{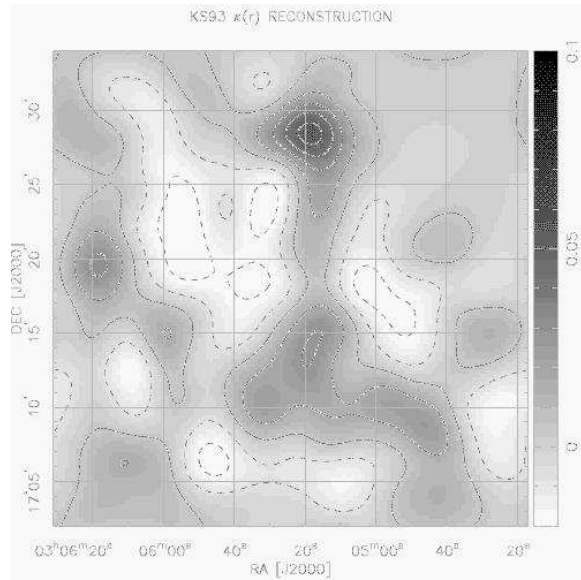}}}
\caption{Mass map \index{mass!map} 
of the supercluster MS0302+17, smoothed on a scale
of 90 arcseconds (black indicates higher density).  Each of the three
densest blobs corresponds to a known galaxy cluster.
From  \cite{Kaiser_supercluster}.}
\label{fig-supercluster}
\end{figure}

Many clusters have now been mapped using these techniques, and the
mass distributions recovered are generally not surprising.  That is,
they roughly follow the optical and X-ray light distributions, on the
scales which weak lensing is able to resolve.  A vivid example of
two-dimensional mass reconstruction is that of the supercluster
MS0302+17 \cite{Kaiser_supercluster} (Figure~\ref{fig-supercluster}).
This supercluster contains three clusters separated by 15-20\arcm\ 
on the sky ($\sim 3-4$ Mpc transverse separation at a
redshift of 0.42).  All three clusters are recovered by the
reconstruction algorithm, above the level of other, presumably noise,
peaks.  This result is robust: it remains when the source selection
criteria are varied, and it disappears when the source positions are
randomized.  There appears to be a filament connecting two of the
clusters, but the authors advise caution, as the signal-to-noise is
low, and real filaments are not expected to have much contrast against all
the other filaments and sheets expected to lie between sources and
observer.  An equally striking reconstruction of the Abell 901/902
supercluster was recently published \cite{Gray2001}.  The close
correspondence of mass peaks and known clusters says something about
the predictability of dark matter, as discussed below.

Despite these successes, it is worth remembering that a map (or radial
profile) of $\kappa$ is not a map of mass.  $\kappa$ can be converted
to mass only with a careful calibration of the source redshifts, which
must include an estimate of source contamination by the cluster
itself.  In massive clusters, magnification provides another source of
error by increasing the mean redshift of sources which have been
selected according to an apparent magnitude cut.  This results in
$\Sigma_{\rm crit}$ \index{density!critical} 
being a function of radius, as it decreases at
small radii where the sources of a fixed magnitude tend to be more
distant \cite{FT97}. While much attention has been paid to optimizing
the formal reconstruction methods, these more mundane problems require
equal attention.


\subsection{Mass and light}
\label{sec-ml}
\index{cluster!mass-to-light ratio}

\begin{figure}
\centerline{\resizebox{3in}{!}{\includegraphics{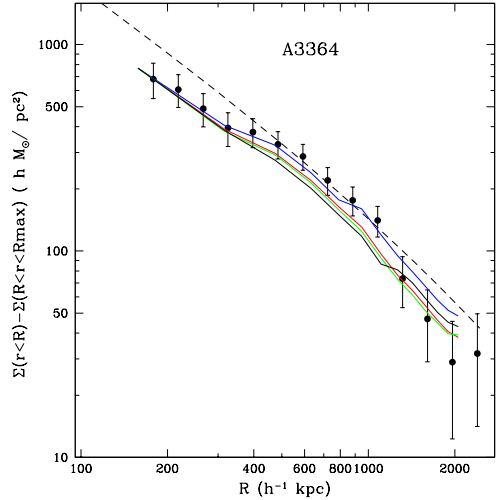}}}
\caption{Projected mass and light density profiles of Abell 3364.
The light profiles were observed in observer-frame
$B_j$ (blue line), $V$ (green), $R$ (red), and $I$ (black) filters,
and computed in the same differential apertures used for the mass.
The light profiles have each been shifted vertically to intersect the
innermost mass point, hence they are in arbitrary units.  Mass follows
light surprisingly well on all measurable scales.  The dotted line
shows the shape of an isothermal profile, which is not quite a
straight line with this estimator, to guide the eye (it has not been
fit to the data).  The two lowest mass points are approaching the
level of systematic error estimated from the point-spread function. 
Note that in the aperture densitometry method, error bars on adjacent
points are not independent, so that the errors should be thought of as
a band. From  \cite{NCS}.}
\label{fig-a3364}
\end{figure}

Does light trace mass ?  The answer must be at least a qualified yes,
because  the projected shapes of cluster lenses tend to
agree with the shapes suggested by their emitted light
(Figure~\ref{fig-a3364}).
However, the qualifications are important !

First, the correspondence holds only on scales larger than
galaxies.  The vast majority of visible-wavelength light from
clusters comes from individual galaxies, not diffuse emission.
Although weak lensing is not well suited to examine small scales,
there is ample evidence from strong lensing that cluster mass
distributions (like their X-ray emissions) do not peak on galaxy
scales \cite{CL0024}.

Second, not all light is equal.  Blue light is dominated by very young
stars, while established stellar populations which presumably trace
mass better tend to be red.  Hence small variations in star formation
could scatter the ratio of mass to blue light $M/L_B$ widely from
system to system, but $M/L_R$ should be much more stable.
Unfortunately, the literature has a tradition of quoting $M/L_B$,
which obfuscates the issue of whether mass is traced by light from
established stellar populations.  Compounding this confusion are
different methods of computing rest-frame emission given the observed
emission, and the occasional quotation of $M/L$ at $z=0$, meaning that
$L$ has first been adjusted to a value that it would have at $z=0$
given passive evolution.  Because stellar populations fade with time,
$L$ can decrease, and $M/L$ increase, significantly from its {\it in
situ} value.
\index{cluster!mass-to-light ratio}
\index{cluster!bias}

Third, even if $M/L$ were more consistently defined, it may not be
constant spatially or as a function of scale.  Although the literature
is full of mass reconstructions which generally follow smoothed light
distributions, the M/L found varies widely, from $\sim 80h$ to $\sim
800h$. Some of this is no doubt due to different methods mentioned
above, but there is reason to believe that not all of it is.  For
example, the high value of $\sim 800h$ for MS1224+20 was found
independently by two different
investigators \cite{Fahlman1994,Fischer1999}. Also, attempts to
uniformly treat samples of $\sim 10$ clusters have found a range of
$M/L$ within the samples \cite{Smail97,NCS}. There are some hints that
some of the scatter may be due to a trend of increasing $M/L$ with
cluster mass.

If so, this follows a broader trend in which groups have lower $M/L$
than the typical cluster \cite{Hoekstra_etal2001}, and typical galaxies  
have still lower $M/L$.  The idea that $M/L$
might depend on environment is called {\it bias}.  Specifically, bias
is when light is more concentrated than mass; the reverse, {\it
antibias}, may also occur.
However, indications of antibias can be understood as a simple
consequence of stellar evolution and the choice of a blue
bandpass \cite{Bahcall2000}. In the end, $M/L$ may be more a question
of star formation history and bandpasses than of the nature of dark
matter. \index{dark matter}

An alternative approach in terms of galaxy-mass correlations may offer
more promise.  Because large mass concentrations are clearly more
associated with early-type galaxies than with later types (the {\it
morphology-density relation}), restricting the analysis to early-type
galaxies might reveal a tighter relationship to mass.  This was first
done for the MS 0302 supercluster, shown in
Figure~\ref{fig-supercluster} \cite{Kaiser_supercluster}.  A
cross-correlation between the projected mass density and that
predicted from the early-type galaxies revealed a strong relationship,
which did not vary with density, whereas a simple $M/L$ would have
acquired variations from the variations in star formation activity.
This approach has been extended to the field, with similar
conclusions \cite{WKL2001}.  Correlations between mass and light were
found to be not so simple in the Abell 901/902
supercluster \cite{Gray2001}, but perhaps the difference is due to all
light, not only that from early-type galaxies, being used in the Abell
901/902 work.  Still, the correlation of mass with early-type galaxies
must fail on some smaller scale, as we know that galaxy groups have
mass but generally no early-type galaxies.  Clearly more work is
required in this area, as the correlation of mass with different types
of emitters may provide clues to the nature of dark matter.

\subsection{Clusters as cosmological probes}
\label{sec-clustprobes}
\index{cluster!mass-to-light ratio}

There is a hidden agenda behind all the effort that has gone into
measuring cluster $M/L$: if cluster $M/L$ is representative of the
universe in general, the mean density of the universe $\Omega_m$ can
be estimated simply by scaling the local luminosity density by this
ratio.  This is one version of the {\it fair sample hypothesis}, and
it is one of the many ways to use clusters as cosmological probes.
These can be divided roughly into methods that extrapolate from
physical conditions in clusters (using clusters only because
they facilitate certain measurements), and methods which use clusters
to diagnose formation of structure in the universe.

Scaling the luminosity density by $M/L$ is the simplest example of the
first type of method.  The fair sample hypothesis remains just a
hypothesis, but it has nevertheless spawned many estimates of
$\Omega_m$, which tend to be
\raisebox{-0.5ex}{$\stackrel{<}{\scriptstyle \sim}$}
0.4 \cite{Smail97,Kaiser_supercluster,NCS,Hoekstra_etal2001}. However,
the apparent variation of $M/L$ with environment and age makes this
approach suspect, and it is worth asking whether any property of
clusters other than light could be used in a similar scaling argument.

The best candidate is the baryon fraction $f_b$, the ratio of baryonic
\index{mass!baryonic}
mass to total mass.  Because there is no reason to believe that infall
into clusters favors baryons over dark matter or vice versa, $f_b$ is
plausibly equal to $\Omega_b / \Omega_m$.  In addition to being
plausible, the baryonic hypothesis is easier to investigate with
simulations of structure formation, because tracking the baryons
is easy in such simulations; the hard part is simulating star
formation and the resultant light emission.
With $\Omega_b$ fairly well known from Big Bang nucleosynthesis
arguments \cite{Burles}, a determination of $f_b$ would quickly yield
$\Omega_m$.  Lensing can provide an estimate of the total mass, while
SZE \index{Sunyaev-Zel'dovich} 
measurements can probe the dominant baryonic component, the
intracluster gas.  Simulations indicate that the combination should
reveal $f_b$ to 10\% or better \cite{Zaroubi,Dore}. The first results
from real clusters (but with total mass estimated from X-ray emission
rather than lensing) indicate $\Omega_m \sim 0.25$ \cite{Grego_etal}.
It should be noted that any census of baryons is likely to be
incomplete, as they can take many forms which are difficult to detect
(brown dwarfs, planets, etc.).  Hence this method provides a lower
limit to $f_b$ and an upper limit to $\Omega_m$.

There is always a chance that physical properties of clusters such as
$f_b$ are simply not representative of the universe in general.  A
second and more powerful class of cosmological probe uses clusters as
tracers of structure.  Only their mass is important, and in
particular, their {\it mass function}, \index{cluster!mass function} 
the number density of clusters as a
function of mass.  The redshift evolution of the cluster mass function
is a probe of $\Omega_m$: all else being equal, a high-density
universe should show more recent evolution than a low-density
universe.  In fact, it has been argued that the existence of even one
massive cluster at high redshift \index{cluster!high redshift} 
({\it e.g.} MS1054 at
$z=0.83$ \cite{Donahue} and now also ClJ1226.9+3332 at
$z=0.88$ \cite{Ebeling}) demonstrates that evolution has not been as
rapid as required if $\Omega_m = 1$ \cite{Bode}. Conclusions based on a
few massive clusters are suspect, however, because as the extreme tail
of a distribution, their numbers are highly dependent on the
assumption of Gaussianity in the primordial fluctuations.  The
argument can even be turned around: given an independent measure of
$\Omega_m$, cluster counts can put strong constraints on primordial
non-Gaussianity \cite{RGS2000,KIS2001}.  With plentiful wide-field data
now available and with weak lensing techniques having been honed on
less massive clusters, it will soon be possible to construct an honest
mass function, \index{cluster!mass function} 
which will constrain both quantities \cite{KS1999}.
The redshift evolution of the cluster mass function can also constrain
dark energy \cite{Hennawi,Huterer}.  Without any uniform weak-lensing
cluster samples, though, we must defer this discussion to Future
Prospects and turn our attention to progress in obtaining such a
sample.

\subsection{Shear-selected clusters}
\label{sec-shearselec}

The use of clusters as cosmological probes centers on clusters as mass
concentrations, not as collections of galaxies and gas.  Yet all cluster
samples compiled to date have been based on emitted light from
galaxies ({\it e.g.}  \cite{ACO}) or from a hot intracluster
medium \cite{xrayreview}.  Because these mechanisms do not involve dark
matter, \index{dark matter} 
which is the dominant component by mass, a mass function based
on these samples may well be biased.  In addition, the $r^{-2}$
falloff of emitted light implies that the high-redshift end of such
samples will always be dominated by the most luminous clusters, which
potentially introduces another bias.  Shear-selected clusters are
needed to investigate these potential biases and provide a clean mass
function, \index{cluster!mass function} 
and this is currently an active area in weak lensing.

%
%
First, a note on terminology.  ``Cluster'' implies a
collection of galaxies, but 
if a large mass concentration with no visible galaxies were to be
identified, it would probably be called a ``dark cluster''.  Although
``dark matter halo'' would be a more accurate term, it is used almost
exclusively in theoretical and computational papers, not observational
work.  Here we shall continue to use the term ``cluster'', but we
emphasize that this is a working hypothesis.  After large samples of
shear-selected mass concentrations are thoroughly followed up with
other methods, it will become clear if a different term is more
appropriate.

Unfortunately, such samples are not available yet.  Although many
previously known clusters were studied with weak lensing in the 1990s,
no surveys for new clusters were conducted, partly because of the
small fields of view afforded by cameras on large telescopes until
later in the decade, and partly because techniques needed to be proven
on known clusters first.  The first serendipitous detections of mass
concentrations came when unexpected peaks appeared some distance from
the target in mass reconstructions of known
clusters.

In the first reported detection, a mass concentration was found
projected near Abell 1942 (7\arcm\ from the center), and confirmed by
a mass map constructed with data from a different camera and at a
different wavelength \cite{Erben2000}.  There is no obvious
concentration of galaxies associated with this mass, although the area
does contain a poor group of galaxies and some weak X-ray emission.
Because the redshift of the mass is unknown, its mass and $M/L$ are
also unknown.  However, with an upper limit on the light in the area,
the lower limit on M/L can be computed as a function of redshift.
There are redshifts for which the object could have a reasonable M/L,
around 400 \cite{Gray2001a}.  This object is therefore not necessarily
more dark than some X-ray selected clusters, which have M/L up to 600
or more \cite{Fahlman1994,Fischer1999} (see Section~\ref{sec-ml}).  If
it is at the redshift of Abell 1942 ($z=0.22$), its M/L is at least 600,
which is very dark but just on the edge of the X-ray selected range,
and perhaps explainable in a merger scenario with Abell 1942 proper.
\index{cluster!mass-to-light ratio}

In a second detection, Hubble Space Telescope (HST) imaging revealed an
extra mass concentration \index{HST} one arcminute from the center of
CL1604+4304 \cite{Umetsu2000}.  It is also seen in a second pointing
shifted by 20\arcs, so it is likely to be real, but the interpretation is
not clear.  It seems likely to be substructure in the cluster rather
than an independent structure, but the necessary followup is lacking.
In a third case, serendipitous mass concentrations were found in a
survey of known clusters \cite{NCS}.  Some of these corresponded to
galaxy groups, which followup spectroscopy showed to be real and not
associated with the target clusters, although in the same general
redshift range (the range to which the lensing survey was of course
most sensitive).  

\begin{figure}
\centerline{\resizebox{5in}{!}{\includegraphics{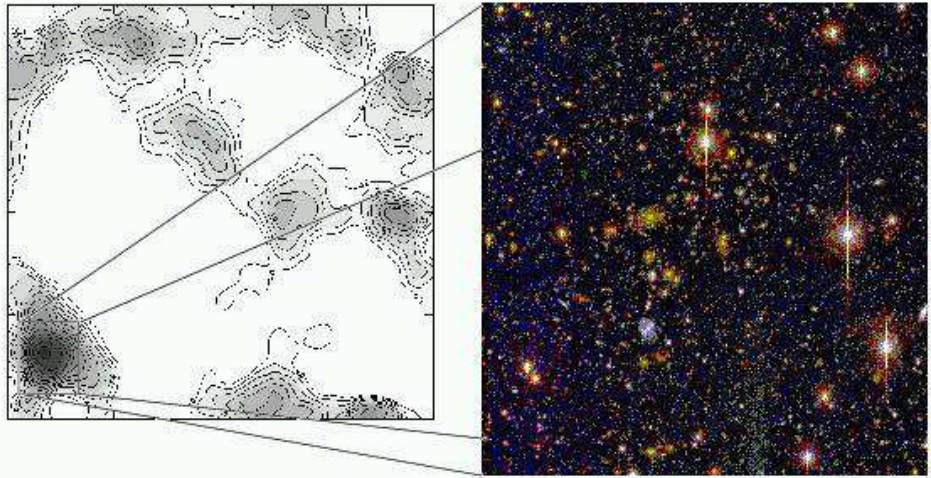}}}
\caption{A shear-selected cluster of galaxies.  At left is a $\kappa$
map of a 40\arcm\ field not selected to contain a previously known
cluster; black indicates higher density. The mass concentration at
lower left corresponds to a cluster of galaxies (inset),
spectroscopically confirmed at a redshift of 0.28 and a velocity
dispersion of 615 km s$^{-1}$.}
\label{fig-shearclust}
\end{figure}

Although these hints were exciting, large ``blank'' fields ({\it i.e.}
fields not selected to contain a known cluster) are more appropriate
for finding unambiguously new clusters, and the first truly convincing
shear-selected cluster was indeed discovered in such a
field \cite{Wittman2001}.  This object is clearly a cluster of galaxies
(Figure~\ref{fig-shearclust}), with a solid spectroscopic redshift
(0.28), velocity dispersion \index{cluster!velocity dispersion} 
(615 km s$^{-1}$), and lens redshift
coinciding with the spectroscopic value (see Section~\ref{sec-tomo}).
The M/L is at the high end of, but definitely within, the range found
for optically and X-ray selected clusters.  Although clearly seen at
visible wavelengths, no X-ray emission is detected at this position.
A second shear-selected cluster has recently been assigned a
spectroscopic redshift \cite{Wittmanetal2002}.  At $z=0.68$, this
cluster begins to fulfill the promise of lensing in terms of avoiding
the $r^{-2}$ falloff of methods which depend on emitted light.  

Finally, the most recent candidate makes perhaps the strongest case
yet for a dark cluster \cite{Mirallesetal2002}.  A tangential alignment
was found around a point in a randomly selected 50\arcs\ STIS field,
significant enough that the data allow only a 0.3\% chance of this
occuring randomly.  There are indications of strong lensing as well.
A nearby group of galaxies could provide enough mass to explain this
only if its M/L is two orders of magnitude higher than expected.  As
with the first two cases cited above, more followup, including a lens
redshift, is desperately needed to make sense of this candidate.

Thus far, the serendipitous shear detections present no clear pattern,
apart from the feeling that these are not typical optically or X-ray
selected clusters.  There is no proof of truly dark clusters,
something which only lensing could detect.  Perhaps this is only for
lack of followup.  Yet, the detection of a truly dark cluster would
reveal surprisingly little about the nature of dark matter.  Rather,
it would indicate that either baryons did not fall into the potential
well created by the dark matter halo, or that star formation failed
there.  These are intriguing scenarios, but they raise questions about
baryons rather than answer questions about dark matter.  

Meanwhile, there are several surveys \index{cluster!survey} 
of tens of square degrees
currently underway \cite{DLS,Descart}, which will yield samples of
dozens of shear-selected clusters, rather than a serendipitous few,
and perhaps yield a better idea of typical and extreme shear-selected
clusters.  Much work remains in terms of settling on estimators which
maximize detection of real mass concentrations while minimizing false
positives.  For example, do we simply look for peaks in convergence
maps (or maps of some other quantity such as potential or aperture
mass), \index{convergence} 
or do we apply a matched filter, which implies that we know
what we are looking for?  While such work has been done theoretically
and computationally \cite{RB99}, we must get our hands dirty with real
samples before we can have much confidence in the scattered examples
published as of today.

%
%
The advantages of shear selection in avoiding baryon and emitted-light
bias are obvious, but no single cluster-finding technique will be
completely unbiased.  SZE selection \cite{Holder} is an exciting new
method which is also independent of emitted light.  This is especially
important in going to high redshift because of the $r^{-2}$ falloff of
emitted light.  In this respect, SZE \index{Sunyaev-Zel'dovich} 
has the advantage because its
background source is at a very high redshift (the cosmic microwave
background at $z \sim 1100$), and because lensing is most efficient at
detecting clusters at much lower redshift than the sources.  However,
lensing and SZE methods are so new that samples are not yet available.
X-ray and optical selection are more established, and X-ray surveys
have recently made great strides in detecting high-redshift
clusters \cite{xrayreview}, indicating that it can compete 
with other methods at \index{cluster!X-ray}
$z>1$ despite the $r^{-2}$ falloff of emitted light.  X-ray emission
has the additional advantage of depending on the square of the local
density, making it less vulnerable to projection effects (although the
density-squared dependence could be viewed as a disadvantage when
trying to determine the total cluster mass).
Table~\ref{table-selection} summarizes the properties of these
selection methods.  In the end, comparison of differently selected
samples will always be necessary, and much work remains to be done
before we can claim that all the important biases are known.

\begin{table}
\centerline{\begin{tabular}{|l|c|c|c|c|} \hline 
Selection & Projection & Emitted & Baryon & Samples\\ 
method & effects? & light?& dependent? & available now?\\ 
\hline
Optical & yes & yes & yes & yes\\
X-ray & no & yes & yes & yes\\
Lensing & yes & no & no & almost \\
SZE & yes & no & yes & almost \\
\hline
\end{tabular}}
\caption{Comparison of cluster selection methods.}
\label{table-selection}
\end{table}

\subsection{Tomography with clusters}
\label{sec-tomo}
\index{tomography}

Judging from the examples of the previous section, followup and
identification of shear-selected clusters will be more difficult than
finding them.  The most basic parameter, redshift, is unknown in
several cases.  Without a redshift, even the lens mass and $M/L$ must
remain unknown, leaving little solid information.  A spectroscopic
redshift is impossible in the case of the Abell 1942 field with no
obvious lens-associated galaxies, and difficult in the CL1604+4304
field, with CL1604+4304 itself projected so nearby.  Thus, there is a
great need for a method of determining the lens redshift from the
lensing information alone.

\begin{figure}
\centerline{\resizebox{2.5in}{!}{\includegraphics{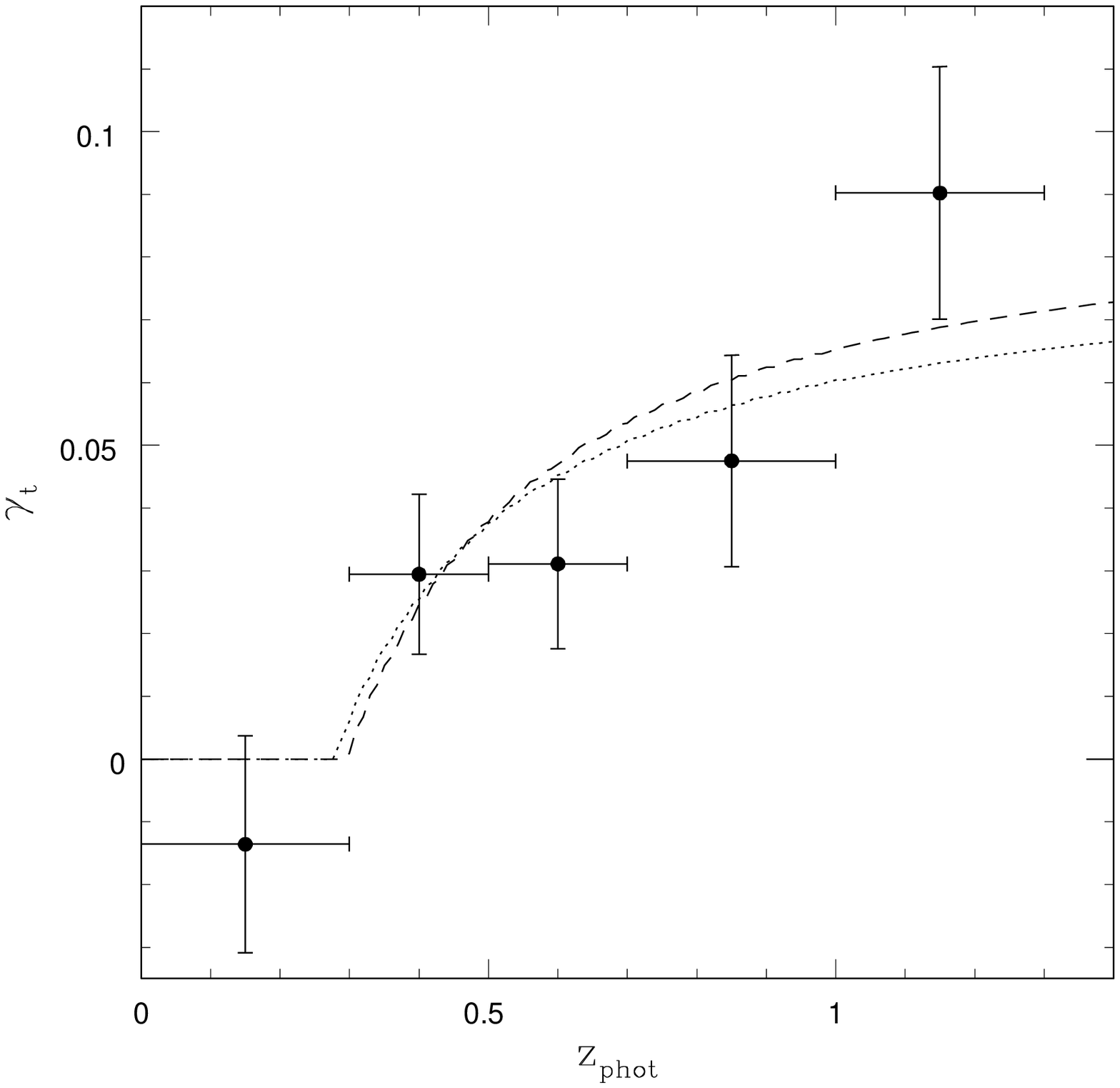}}
\resizebox{2.3in}{!}{\includegraphics{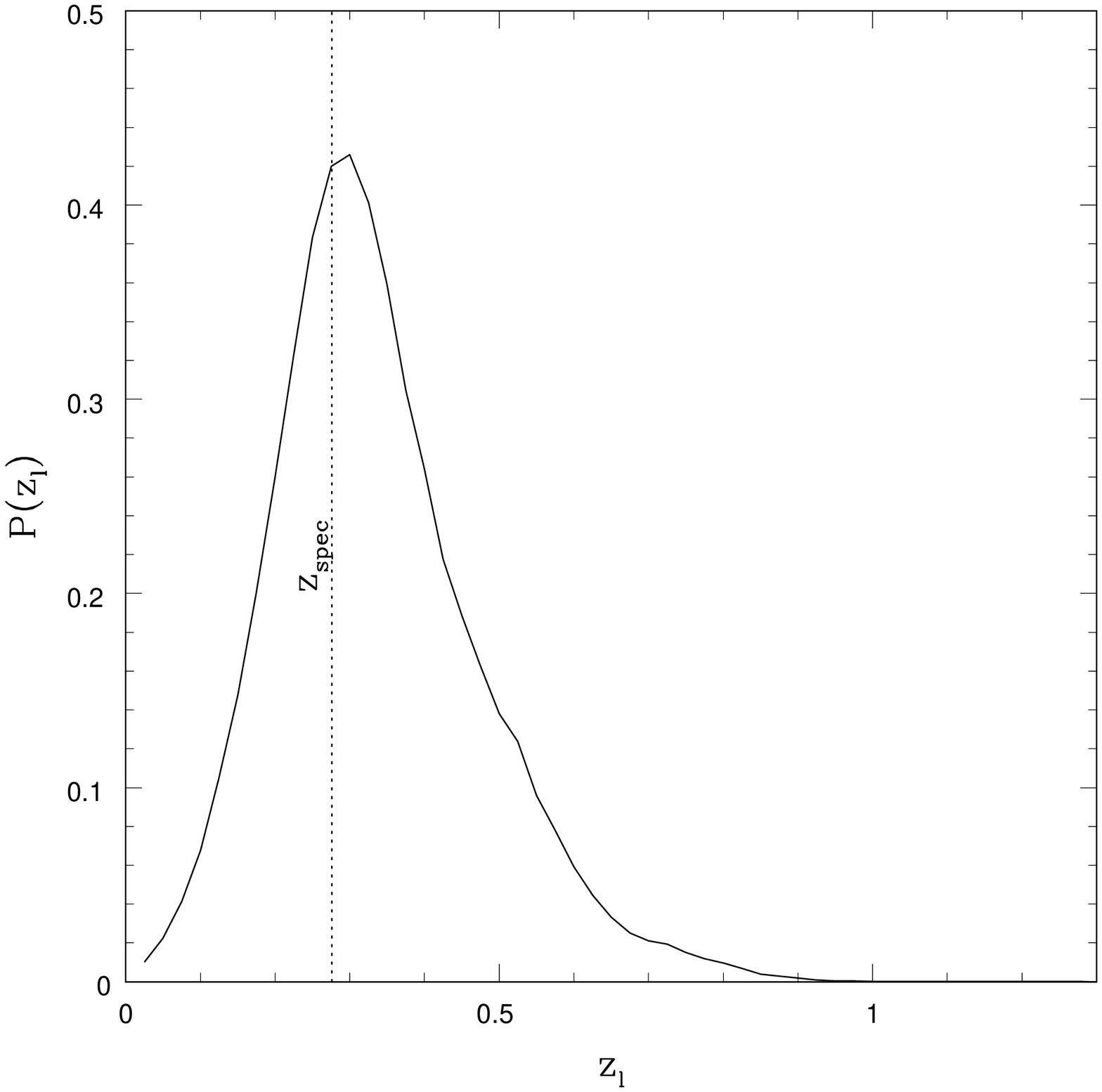}}
}
\caption{{\it Left}: tangential shear as a function of source
(photometric) redshift. The dotted line is the best fit for a lens at
the cluster spectroscopic redshift of 0.28, while the dashed line is
the best fit with the lens redshift derived from the lensing data
alone.  ($z=0.30$).  {\it Right}: The lens redshift probability
distribution derived from the data at left.  The method is promising:
The most likely lens redshift is within 0.03 of the cluster
spectroscopic redshift, but the width of the distribution is $\sim
0.1$, indicating the need for more precise data.  From
 \cite{Wittman2001}.}
\label{fig-tomo}
\end{figure}

If sources can be differentiated by redshift, the redshift of a lens
will be revealed by the way that shear increases with source redshift
(Figure~\ref{fig-dratio}).  Photometric redshifts 
\index{photometric redshift} are required for the
sources, but this is straightforward if the deep imaging required for
the shear measurement is extended to multiple filters.  Two-filter
imaging is routinely done anyway, to filter the sources based on
color.  Four filters is sufficient to provide photometric redshifts
accurate to $\sim 0.1$ on each source, which is accurate enough given
the large amount of shape noise on each galaxy and the breadth of the
lensing kernel.  This method has been demonstrated on one
cluster \cite{Wittman2001} (Figure~\ref{fig-tomo}).  The most likely
lens redshift is within 0.03 of the spectroscopic redshift ($z=0.28$), but
the formal error estimate is $\sim 0.1$.

Obviously, lens redshifts cannot compete with cluster spectroscopic,
or even photometric, redshifts with this level of precision.  Some
improvement is to be expected as photometric redshifts improve.  For
example, the filter set used was not designed to be optimal for
photometric redshifts, but future large surveys will be paying close
attention to this issue.  Also, the work cited neglected to use
sources which were undetected in one or more filters, but photometric
redshifts are not impossible to assign to such sources, and their
inclusion could improve the statistics.  A rigorous treatment would
take account of each source's photometric redshift error estimate, and
so on.  Work is needed on optimal lens redshift algorithms.

Still, lens redshifts can be useful even at this level of accuracy.
The most basic use is to confirm the unstated assumption in all
cluster weak lensing work to date---that the cluster {\it is} the lens, not
merely in the same line of sight.  While no doubt valid, confirmation
of such basic assumptions is always welcome.  Second, in cases where
dark mass concentrations are found without any associated galaxies, a
lens redshift is the only way to constrain the basic parameters of the
mass concentration.  Indeed, if there is any skepticism about such
claims, it would be conclusively dispelled by demonstrating that the
observed shear increases with source redshift in the predicted way.
Third, large weak lensing surveys may find enough shear-selected
clusters to make complete spectroscopic followup burdensome.  In that
case, rough lens redshifts may be good enough for examining statistics
of many clusters, or at least for identifying the more interesting
candidates for followup.  Therefore, this type of tomography 
\index{tomography} will
probably be a routine feature of future shear-selected surveys.

Finally, note how the spread in source redshifts would have caused
more uncertainty in shear had source redshifts not been known in
Figure~\ref{fig-tomo}.  One way of improving cluster shear
measurements, which until now have used at most a color cut to avoid
contamination of the sources by cluster members, will be the use of
photometric redshifts to weight sources.  If Figure~\ref{fig-tomo} is
any guide, this might make an improvement of up to a factor of two.

\section{Large-scale structure}
\index{large-scale structure}
 
Clusters are not the largest structures in the universe.  Although it
had long been known that clusters themselves tend to cluster, it was
only in the 1980's that redshift surveys began to reveal apparently
coherent structures---filaments and voids---on very large scales, up
to $\sim 50$ Mpc.  Current redshift surveys are producing impressive
views of this foamy galaxy distribution out to $\sim 600$ Mpc, or $z
\sim 0.2$  \cite{2dF}.  But what about the mass distribution ?

\begin{figure}
\centerline{
\resizebox{4in}{!}{\rotatebox{-90}{\includegraphics{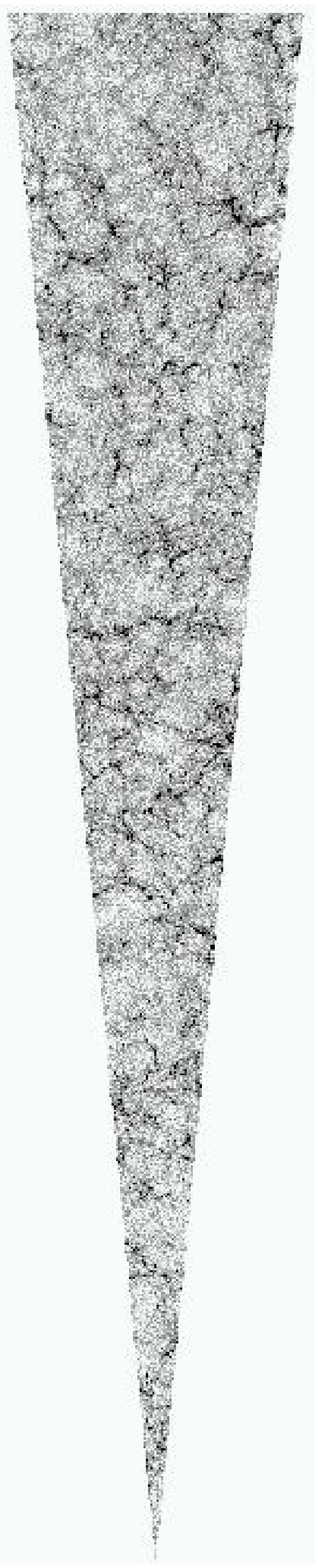}}}}
\centerline{
\resizebox{4in}{!}{\rotatebox{-90}{\includegraphics{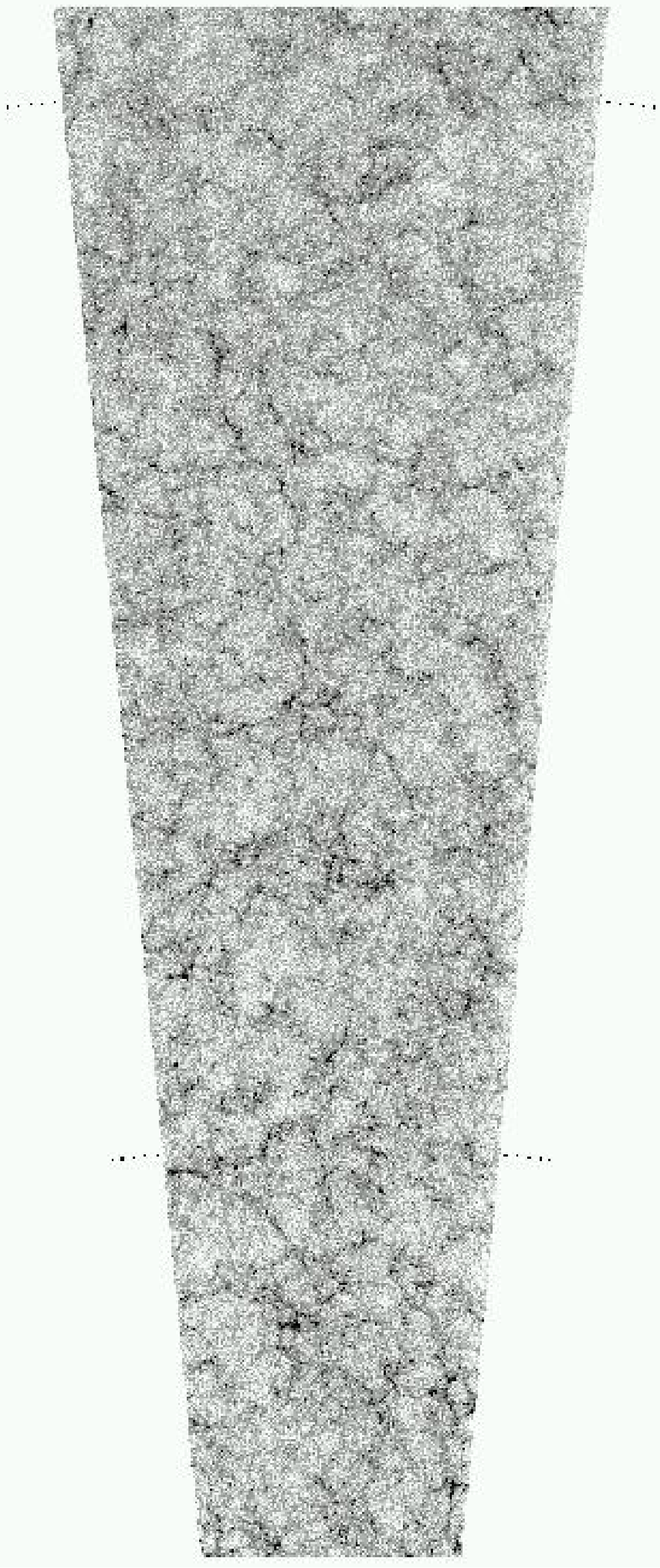}}}}
\caption{Simulation from the Virgo collaboration showing the evolution
of large-scale structure in a 7$^\circ$ slice of a $\Lambda$-dominated
universe, with black indicating highest density.  The dotted lines
indicate $z=1$ and $z=2$. Adapted from  \cite{Virgo}. }
\label{fig-lss}
\end{figure}

\index{redshift surveys}
\index{cosmology}
Simulations of cold dark matter show similar structures in mass
(Figure~\ref{fig-lss}).  Furthermore, they show how the evolution of
large-scale structure depends on cosmological parameters and on the
nature of dark matter.  Good measurements of large-scale structure
evolution should therefore be able to constrain cosmological
parameters and the nature of dark matter through comparison with
simulations.  Weak lensing is a good candidate for such comparisons,
because like the simulations it deals with mass, not galaxies; and
because it can easily reach back to $z \sim 1-2$, providing a long
baseline in cosmic time.

Figure~\ref{fig-lss} illustrates just how many voids and filaments 
\index{mass!filaments} are
expected to lie between us and a source at $z \sim 1$. Because of
projection effects, weak lensing will never produce stunningly
detailed three-dimensional mass maps to allow comparisons with such
simulations.  But weak lensing by large-scale structure does leave a
statistical signature.  This ``cosmic shear'' is a strong function of
cosmological parameters, in particular, $\Omega_m$ and $\sigma_8$, the
rms density variation on 8 Mpc scales, and thus is potentially a very
useful cosmological tool.  However, cosmic shear leaves a much weaker
signal than do clusters, making detection more difficult and
systematics more dangerous.  The first detections of cosmic shear came
in 2000, a decade after weak lensing by clusters was detected.  For
that reason, cosmic shear is just beginning to take its place in the
cosmology toolbox.

\subsection{Cosmic shear estimators}
\label{sec-cosmicshearstats}
\index{shear!cosmic} \index{cosmology}
Cosmic shear, unlike the shear induced by clusters and groups, has no
center, and that has led to the formulation of a variety of statistics
different from those used to analyze clusters.  We summarize them here
to provide the basis for interpreting the results presented here and
in the literature.  Each of the following statistics has advantages
and disadvantages, and current practice is to report results in terms of
several different estimators to verify robustness.

A few comments apply to all the estimators mentioned below.  Current
wide-field cameras have fields of view of $\sim 0.5^\circ$, and all
results to date have been reported on these or smaller scales.  But a
look at Figure~\ref{fig-lss}, with its opening angle of 7$^\circ$,
shows that such small fields will give different results depending on
where they happen to lie.  Because of this {\it cosmic variance}, one
such field cannot really constrain the cosmology.  A sample of
randomly chosen fields is required, with the field-to-field scatter in
results giving some idea of the cosmic variance.  Observed variance
could also be due to problems with the instrument or telescope, so
this is really an upper limit to the cosmic variance, but it is still
a very useful number.  Some groups are currently doing much larger
fields by stitching together multiple pointings, sometimes with sparse
sampling, but multiple, widely separated fields are still required to
insure that cosmic variance has been beaten.

On small scales, the dominant statistical noise source is simply shape
noise, but systematics are also larger here.  Small-scale PSF
variations cannot be mapped because the density of stars is too low;
intrinsic alignments play a larger role on small scales; and
comparison to theory (not necessarily simulations) is hampered by the
difficulty of modeling the nonlinear collapse of dense regions.
Results on scales $< 1$\arcm\ may say more about nonlinear collapse
and possibly intrinsic alignments than about the cosmology.


\subsubsection{Mean shear}
\index{shear!mean}
The mean shear in a field (simply averaging all sources) will in
general be nonzero in the presence of lensing.  However, it will tend
to zero for a field of any significant size, so this statistic is of
limited use.  We mention it for completeness, as some early work with
small fields of view used this statistic.  But mean shear in a small
field of view is difficult to interpret, as it could result from a
single structure projected near the line of sight.

\subsubsection{Shear variance}
\index{shear!variance} 
The next logical step is to compute the variance, among a group of
boxes of angular size $\phi$, of the mean shear in each box.  
Because variance is a positive definite quantity, noise rectification must be
subtracted off.  
The importance of the noise rectification term depends on the number
of sources per box.  Roughly speaking, for $\phi<1$\arcm, the noise
rectification term is larger than the lensing signal itself, but for
larger $\phi$ the lensing contribution dominates and at several
arcminutes the noise correction becomes quite small.  Thus, measurements of
shear variance at $\phi<1$\arcm\ should be treated with caution (in
addition to the cautions cited above for small scales).  Usually the
results are presented as a function of $\phi$, but the values for
different $\phi$ have been computed from the same data.  Hence the
errors on the different scales are not independent, and results are
sometimes less significant than appears at first glance.  The true
significance of such results can be explored with bootstrap
resampling.  A rule of thumb suggested by bootstrap tests is that
measurements on widely differing scales (factors of 10) are largely
independent of each other even when computed from the same data.

\subsubsection{Ellipticity correlations}
\index{ellipticity!correlation}
\index{source!ellipticity}

The observed ellipticities of lensed sources are correlated, so it is
natural to construct a correlation function which measures this effect
as a function of angular separation between sources.  A simple
correlation of the ellipticity components $e_+$ and $e_\times$ would have
little physical meaning, though, as they depend on the orientation of
the detector axes.  If the components of a pair of galaxies are
instead defined with respect to an imaginary line joining their
centers, their correlation does have a physical
significance \cite{Jordi1991b}.
In fact, three useful functions can be defined:
\begin{eqnarray}
\xi_1(\theta) \equiv \langle e_+^i e_+^j\rangle \nonumber \\
\xi_2(\theta) \equiv \langle e_\times^i e_\times^j\rangle \\
\xi_3(\theta) \equiv \langle e_+^i e_\times^j\rangle \nonumber
\end{eqnarray}
where superscripts label the sources and brackets denote averaging over
all pairs of galaxies $i \ne j$ with angular separation $\theta$.

Like shear variance, lensing induces $\xi_1 > 0$ for all $\theta$, but
decreasing with $\theta$.  Unlike shear variance, the computation of
$\xi_1$ does {\it not} result in a positive definite quantity, so
spurious results may be easier to identify.
The behavior of $\xi_2$ in the presence of lensing by large-scale
structure is more interesting: at $\theta = 0$, $\xi_2$ and $\xi_1$
are equal, but $\xi_2$ drops more rapidly and goes negative at some
$\theta$ which depends on the cosmology ($\sim 0.5-1^\circ$). $\xi_3$
is a control statistic.  Unaffected by lensing, it should vanish in
the absence of systematic errors or intrinsic alignments (this is
equivalent to rotating one of each pair of galaxies by 45$^\circ$).
Taken together, these properties provide a signature with several
lines of defense against systematic error.

Like shear variance, values for different $\theta$ are computed from
the same data, so the same warnings about nonindependent angular bins
apply.  Unlike shear variance, though, there are two independent
quantities ($\xi_1$ and $\xi_2$) at each $\theta$, which can be
checked against each other.  For example, unless $\xi_1(0) = \xi_2(0)$
and $\xi_1(\theta) > \xi_2(\theta)$ for $\theta > 0$, the results are
suspect.  If these checks make sense, $\xi_1$ and $\xi_2$ can be
combined into a single higher signal-to-noise measurement.  In fact,
shear variance can be understood as $\xi_1(0) + \xi_2(0)$, convolved
with a square window function of width $\phi$.
\index{shear!variance}

\subsubsection{Aperture mass}
\index{mass!aperture}

The aperture mass statistic $M_{ap}$ was designed to address the
nonlocality of shear.  It is a generalization of the $\zeta$ statistic
already mentioned in the context of clusters \cite{Fahlman1994,KSB}.
As in the $\zeta$ statistic, tangential shear is computed in a
circular aperture, but here it is weighted with a compensated filter
function; the weight is positive at the center of the aperture and
negative at the edges, for a total weight of zero.  This has the
remarkable property of making adjacent apertures nearly independent,
whereas the shear in adjacent apertures is highly correlated.
$M_{ap}$ was first suggested as a way of looking for clusters in
wide-field images \cite{Schneider1996} , and later its variance
$\langle M_{ap}^2\rangle$ was proposed as a measure for cosmic
%
%
shear \cite{Schneider_apmass} (note that because the total weight
vanishes, so does the expectation value: $\langle M_{ap}\rangle = 0$).
Although aperture mass tends to be noisier than the other estimators,
its compensating virtue is that measurements on different scales are
almost completely independent.

\subsubsection{Other estimators}
All of the above estimators are at most two-point statistics.
Higher-order statistics have been proposed.  For example, values of
the projected mass field ($\kappa$ or $M_{ap}$) should have a skewness
due to many somewhat underdense regions (voids) and a few extreme
overdense regions.  This skewness depends on $\Omega_m$ and the matter
power spectrum in a different way than does the variance, leading to
suggestions that together they could constrain both
quantities \cite{Bernardeau97}.  This non-Gaussianity may also be
revealed through morphological analysis of convergence
fields \cite{Sato,Taruya,Guimaraes}.  All this remains largely theoretical,
as these high-order statistics are in practice noisier than the
two-point estimators which are providing the first detections of
cosmic shear. \index{shear!cosmic} 
There is a recent claim of detection of this
non-Gaussian signature \cite{skew}, but an accurate measurement of
skewness requires that rare massive halos be present in the
sample \cite{CH2001}, hence a very large area is required.  The best
way to the power spectrum itself may be through maximum likelihood
fitting to the shear data \cite{HW2000,CH2000}.

\subsection{Observational status}
\index{weak lensing!observations}
 
Although the idea of weak lensing by large-scale structure was first
suggested in the 1960's \cite{KS66,Gunn67}, the effect escaped
detection for over three decades.  The first attempts at detection
gave null results \cite{Kristian67,VJT83}, which is not surprising
given the subtleness of the effect ($\sim 1\%$ shear) and the lack of
sensitivity and nonlinearity of photographic plates.
The first analysis of CCD data, albeit with the narrow field afforded
by CCDs in 1994, also yielded only upper limits \cite{Mould}.
With the advent of large-format CCD mosaics, detection was inevitable,
and four groups \cite{WTKDB,vW2000,BRE,KWL} announced detections in
the span of one month in 2000.  

\begin{figure}
\centerline{\resizebox{3in}{!}{\rotatebox{-90}{\includegraphics{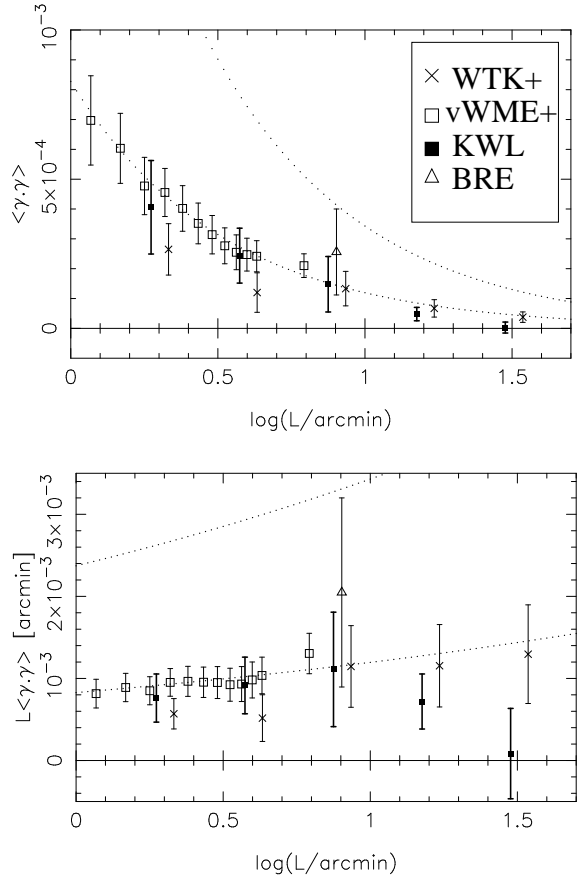}}}}
\caption{First detections of cosmic shear, \index{shear!cosmic} 
in terms of shear variance
versus angular scale.  \index{shear!variance} 
The results of four different groups using
three different telescopes and four different cameras are shown, with
good agreement.  Note that different angular bins from the same
experiment are not independent.  The dotted lines are for two
different source \index{source!redshift distribution} 
redshift distributions (lower, $\langle z \rangle =
1$; upper, $\langle z \rangle = 2$) in a $\Lambda$CDM universe.
Adapted from  \cite{KWL}.  }
\label{fig-cosmicshear1}
\end{figure}

Their results are summarized in terms of shear variance in
Figure~\ref{fig-cosmicshear1}.  The groups used four different cameras
on three different telescopes, with different observed bandpasses and
data reduction procedures and analysis techniques, yet the results
were in good agreement.  This has been taken as proof that
instrumental effects and systematic errors have been vanquished, but
in fact, the results should {\it not} agree if the data and source
selection resulted in different mean source redshifts.  The fact that
the areal density of sources used was similar for all four groups
suggests that the source redshifts were similar despite the different
approaches.  But the possibility remains that different source
redshifts are hiding some disagreement.

Nevertheless, all results point to a low-$\Omega_m$ universe.
Figure~\ref{fig-cosmicshear1} shows the good fit to $\Lambda$CDM.  It
is difficult to constrain $\Lambda$ with these measurements, but shear
variance should scale roughly with $\Omega_m$, so it is clear that
$\Omega_m = 1$, for example, is ruled out.  While this was no
surprise, it signaled the emergence of cosmic shear as a new way to
constrain $\Omega_m$, completely independent of traditional methods
(supernovae, CMB, age of the oldest stars in conjunction with the
Hubble constant, etc.).

Since then, there have been further detections both in
ground-based \cite{Maoli,HYG2001} and
space-based data \cite{Rhodes}.
The state of the art is a many-sigma detection (whatever estimator is
chosen) over 6.5 deg$^2$ leading to quantitative constraints in the
$\Omega_m,\sigma_8$ plane \cite{vW2001}.
There is a significant degeneracy between $\Omega_m$ and
$\sigma_8$; the first generation of cosmic shear papers simply assumed
a value of $\sigma_8$ consistent with the local abundance of clusters.
At the same time, efforts to improve the signal-to-noise of cosmic
shear measurements by decomposition of cosmic shear into E and B modes
are underway and appear to have met with success \cite{modes}.
 

Currently, several large (tens of square degrees) surveys are under
way with the goal of very high signal-to-noise analyses of cosmic
shear \cite{DLS,Descart,HYG2001}.  At the same time, the question of
how accurate the measurements can ultimately get is being
explored \cite{Bacon2000,Erben2001,Kuijken}.
However, a word of caution is in order.  Such analyses tend to ignore
the fact that the source redshift distribution is not well known.  The
putative accuracy of current and near-future cosmic shear measurements thus
tends to be far too optimistic.  For the moment, the most accurate
measurements in an absolute sense will be those which are no deeper
than current redshift surveys.
Within a few years, though, this will probably change as photometric
redshifts are used to estimate the source redshift distribution
accurately enough.  Photometric redshifts, in fact, will enable
probing the redshift evolution of cosmic shear; division of sources
into just two or three redshift bins can greatly improve the
measurements of cosmological parameters, specifically $\Omega_\Lambda$ by a
factor of $\sim 7$  \cite{Hu1999}.

\section{Future prospects}

\subsection{New applications}

It is impossible to predict what new applications weak lensing might
find, but it is worth discussing one example of an interesting new
direction: constraints on theories of gravity.  It is unlikely that
weak lensing will serendipitously reveal some new feature of gravity,
because the lenses through which we look are not well calibrated.  But
given an alternative theory of gravity, we can ask if weak lensing
observations are consistent with other observations.
\index{gravity!theory}

Modified gravity is an attempt to explain differences between light
distributions and inferred mass distributions without invoking dark
matter. \index{dark matter} 
It is possible to modify Newtonian gravity to account for
some of the observed differences such as flat rotation curves in
galaxies, but a general correlation between mass and light remains.
If lensing were to find severe discrepancies between mass and light,
such as a dark cluster \index{cluster!dark} 
or clear misalignment of cluster mass and light
axes, this would represent a serious blow to modified
gravity \cite{Sellwood}.  There are some promising dark cluster
candidates (Section~\ref{sec-shearselec}), and Abell 901b presents
mass and light axes which apparently differ \cite{Gray2001}, but there are
no bulletproof examples.  Weak lensing surveys of significant areas
are only now underway, so it will take some time before dark clusters
can be ruled in or out with much confidence.  Note that dark clusters
are not {\it expected} in dark matter scenarios; mass concentrations
should accumulate enough baryons to become visible, if only in X-rays.
Thus an absence of dark clusters would not favor modified gravity over
dark matter, but their presence of would disprove modified gravity as
currently envisioned. \index{cluster!dark}

Recently, the first quantitative predictions of weak lensing in
modified gravity scenarios were published.  Modified gravity, by
increasing the strength of gravity on large scales, would greatly
enhance cosmic shear, inconsistent with measurements.  Thus, at the
large scales probed by cosmic shear, the $r^{-2}$ force law cannot be
modified---if gravity does depart from $r^{-2}$, it is only on scales
from 10 $h^{-1}$ kpc to 1 $h^{-1}$ Mpc \cite{WhiteKochanek2001}.  Weak
(and strong) lensing can also address modified gravity by constraining
halo flattening \cite{MortlockTurner2001}.  Weak lensing by large-scale
structure can also provide a test of higher-dimensional
gravity \cite{UzanBernardeau2001}. \index{gravity!theory}

\subsection{New instruments}
\index{weak lensing!observations}

Today's surveys of tens of square degrees will take years to find
perhaps dozens of shear-selected clusters and put some constraints on
$w$, the dark energy \index{dark energy} 
equation of state, as well as $\Omega_m$
 \cite{Hennawi}.  Tight constraints on both would require a very deep
survey of 1000 deg$^2$  \cite{Huterer}, taking decades with current
telescopes and instruments.  The latest generation of 8-m class
telescopes does not really help, as their fields of view are small,
typically $\sim 10$\arcm\ or less (an exception is the Subaru
telescope which has a $\sim 24$\arcm\ field of view with its SuPrime
camera, a likely source of weak-lensing results in the near future).
A new generation of wide-field telescopes specifically designed for
surveys will dramatically accelerate our ability to do cosmology with
large lensing surveys.  These surveys will cover an area comparable to
that of SDSS, \index{SDSS} but much more deeply.  
The first of these to be funded
is VISTA, \index{VISTA} 
a 4-m telescope with a 1$^\circ$ field of view currently in
the design stage, but apparently it will be infrared-only, limiting
its usefulness for weak lensing.  LSST, an 8-m class telescope with a
3$^\circ$ field of view \cite{LSST} and concentrating on visible
wavelengths, may also be built within a decade.

\begin{figure}
\centerline{\resizebox{3in}{!}{\includegraphics{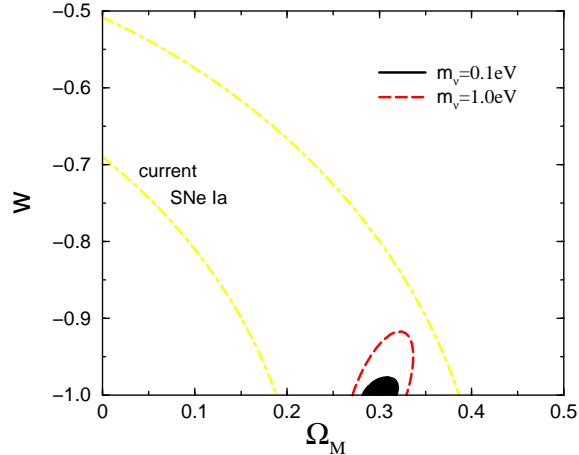}}}
\caption{ 68\% confidence limit constraints on $\Omega_M$ and $w$ for
two values of $m_\nu$, for a weak lensing survey of 1000 deg$^2$ down
to $R=27$, with photometric redshifts providing the source redshift
distribution.  Current 1-$\sigma$ constraints from type Ia supernovae
are shown for comparison. From  \cite{Huterer}.}
\label{fig-huterer}
\end{figure}

Figure~\ref{fig-huterer} shows the potential of a 1000 deg$^2$ survey
which LSST could easily accomplish.  Of course, predictions such as
these depend on the extrapolation of $\sqrt{n}$ statistics to extremely
large areas, so it is wise to ask what systematic effects might
provide a higher noise floor.  Early work on cosmology constraints
from cluster counts assumed NFW profiles \index{mass!profile} for all
clusters \cite{KS1999}.  It was then realized that the profile makes a
big difference, so that cluster counts may tell us more about
dark-matter profiles than about \index{cosmology} 
cosmology \cite{BKS2001}.  However, new
estimators have been proposed to circumvent this
problem \cite{Hennawi}.  Careful attention must also be paid to the
issue of completeness versus false positives in cluster-detection
surveys \cite{White2001}.  Still, by the
time LSST starts operation, these issues may be worked out, and it may
be well to survey all the sky visible from the site.  Such a survey
would also provide a shear \index{shear!power spectrum} 
power spectrum comparable in accuracy to the 
CMB power spectra of today.

Another probe of cosmology which may become feasible with such massive
surveys involves the angular power spectrum of clusters.  The linear
part of this power spectrum is essentially a standard ruler calibrated
by the CMB, so that a power spectrum of clusters at a particular
redshift yields the angular diameter distance to that redshift.  A
very large survey ($\sim 4000$ deg$^2$) could determine this as a
function of redshift, which of course would yield an absolute calibration
of the distance scale and the Hubble constant \cite{CHHJ2001}.
\index{Hubble constant} 

\subsection{New algorithms}

The combination of lensing data with other types of data has been an
active theoretical area recently, and some of these algorithms will
soon prove themselves observationally.  Lensing plus SZE measurements
of clusters will reveal the baryon fraction in that environment,
perhaps leading to a new estimate of $\Omega_m$ from baryon scaling
arguments---or perhaps leading to new aspects of cluster formation.
Combinations of lensing and SZE plus X-ray data will help deproject
cluster mass and gas distributions \cite{Zaroubi}.

Cross-correlation of lensing by large-scale structure with the CMB
\index{CMB} 
will reveal parameters largely hidden from traditional CMB analyses,
such as dark energy, the end of the dark ages, and the gravitational
wave amplitude \cite{Hu2001a,Hu2001b}.  Lensing of the CMB itself may
be detectable by the Planck satellite and constrain the amplitude of
mass fluctuations between us and $z\sim 1000$  \cite{Takada2000}, but
this may have to wait for even higher-sensitivity CMB
probes \cite{Hu2000}.



\begin{thebibliography}{100}
\addcontentsline{toc}{section}{References}

\bibitem{ASW1998} H. Abdelsalam, P. Saha, L. Williams: AJ
\textbf{116}, 1541 (1998)

\bibitem{ACO} G. Abell, H. Corwin, R. Olowin: ApJS \textbf{70}, 1 (1989)

\bibitem{Allen1998} S. Allen: MNRAS \textbf{296}, 392 (1998)

\bibitem{ASF2001} S.~W. Allen, R.~W. Schmidt,
A.~C. Fabian: MNRAS \textbf{328}, L37 (2001)

\bibitem{Bacon2000} D. Bacon, A. Refregier, D. Clowe, R. Ellis: MNRAS \textbf{325}, 1065 (2001)

\bibitem{BRE} D. Bacon, A. Refregier, R. Ellis: MNRAS \textbf{318},
625 (2000)

\bibitem{Bahcall2000} N. Bahcall, R. Cen, R. Dav{\' e}, J. Ostriker, 
Q. Yu: ApJ \textbf{541}, 1 (2000)

\bibitem{BB97} C. Balland, A. Blanchard:
ApJ \textbf{487}, 33 (1997)

\bibitem{BKS2001} M. Bartelmann, L. King, P. Schneider: A\&A \textbf{378}, 361 (2001)

\bibitem{Bartelmann95} M. Bartelmann: A\&A \textbf{303}, 643 (1995)

\bibitem{BN95} M. Bartelmann, R. Narayan: ApJ \textbf{451}, 60 (1995)

\bibitem{Bartelmann1996} M. Bartelmann, R. Narayan, S. Seitz,
P. Schneider: ApJ \textbf{473}, 610 (1996)

\bibitem{skew} F. Bernardeau, Y. Mellier, L. van Waerbeke:  ApJL,
submitted, astro-ph/0201032 (2002)

\bibitem{Bernardeau97} F. Bernardeau, L. van Waerbeke,  Y. Mellier:  A\&A
\textbf{322}, 1 (1997)

\bibitem{BJ2001} G. Bernstein, M. Jarvis: AJ \textbf{123}, 583 (2002)

\bibitem{Biviano2001} A. Biviano: to appear in "Tracing Cosmic Evolution
with Galaxy Clusters", ASP Conference Series; astro-ph/0110053 (2001)

\bibitem{Blakeslee} J. Blakeslee: astro-ph/0108253 (2001)

\bibitem{Blandford91} R. Blandford, A. Saust, T. Brainerd,
J. Villumsen: MNRAS \textbf{251}, 600 (1991)

\bibitem{Bode} P. Bode, N. Bahcall, E. Ford, J. Ostriker: ApJ
\textbf{551}, 15 (2001)

\bibitem{xrayreview} S. Borgani, L. Guzzo: Nature \textbf{409}, 39 (2001)

\bibitem{Brainerd99} T. Brainerd, C. Wright, D. Goldberg,
J. Villumsen: ApJ \textbf{ 524}, 9 (1999)

\bibitem{BTP95} T. Broadhurst, A. Taylor, J. Peacock: ApJ \textbf{438}, 49
(1995)

\bibitem{BTHD2000} M. Brown, A. Taylor, N. Hambly, S. Dye: MNRAS, submitted,
astro-ph/0009499 (2000)

\bibitem{Burles} S. Burles, K. Nollett, M. Turner: ApJL \textbf{552}, L1
(2001)

\bibitem{CKD2000} P. Catelan, M. Kamionkowski, R. Blandford: MNRAS,
\textbf{320L}, 7 (2001)

\bibitem{Cen97} R. Cen: ApJ \textbf{485}, 39 (1997)

\bibitem{shapelets3} T.-C. Chang, A. Refregier: ApJ, \textbf{570}, 447 (2002)

\bibitem{CS2001} D. Clowe, P. Schneider: A\&A \textbf{379}, 384 (2001)

\bibitem{redshiftsurvey} J. Cohen, D. Hogg, R. Blandford, L. Cowie, 
E. Hu, A. Songaila, P. Shopbell, K. Richberg: ApJ \textbf{538}, 29
(2000)

\bibitem{Connolly1995} A. Connolly, I. Csabai, A. Szalay, D. Koo,
R. Kron, J. Munn: AJ \textbf{110}, 2655 (1995)

\bibitem{CH2000} A. Cooray, W. Hu: ApJ \textbf{554}, 56 (2001)

\bibitem{CH2001} A. Cooray, W. Hu: ApJ \textbf{548}, 7 (2001)

\bibitem{CHHJ2001} A. Cooray, W. Hu, D. Huterer, M. Joffre: ApJL \textbf{557},
7 (2001) 

\bibitem{CNPT2000a} R. Crittenden, P. Natarajan, U. Pen, T. Theuns:
ApJ \textbf{559}, 552 (2001)
\bibitem{CNPT2000b} R. Crittenden, P. Natarajan, U. Pen, T. Theuns:
ApJ \textbf{568}, 20 (2002)
\bibitem{CM2000} R. Croft, C. Metzler: ApJ \textbf{545}, 561 (2000)

\bibitem{Descart} http://terapix.iap.fr/Descart

\bibitem{Donahue} M. Donahue, G. Voit, I. Gioia, G. Luppino,
J. Hughes, J. Stocke: ApJ \textbf{502}, 550 (1998)

\bibitem{Dore} O. Dore, F. Bouchet, Y. Mellier, R. Teyssierm:
A\&A \textbf{375}, 14 (2001)

\bibitem{D2001} S. Dye, A.N. Taylor, T.R. Greve, O.E. Rognvaldsson,
E. van Kampen, P. Jakobsson, V.S. Sigmundsson, E.H. Gudmundsson, J.
Hjorth:  A\&A \textbf{386}, 12 (2002)

\bibitem{Ebeling} H. Ebeling, L.R. Jones, B.W. Fairley, E. Perlman,
C. Scharf, D. Horner: ApJL \textbf{548}, 23 (2001)

\bibitem{Erben2000} T. Erben, L. van Waerbeke, Y. Mellier,
P. Schneider, J.-C. Cuillandre, F.J. Castander, M. Dantel-Fort: A\&A 
\textbf{355}, 23 (2000)

\bibitem{Erben2001} T. Erben, L. van Waerbeke, E. Bertin, Y. Mellier,
P. Schneider: A\&A \textbf{366}, 717 (2001)

\bibitem{FGS85} E. Falco, M. Gorenstein, I. Shapiro: ApJ \textbf{289}, L1 (1985)

\bibitem{Fahlman1994} Fahlman, G., Kaiser, N., Squires, G., \& Woods,
D.: ApJ \textbf{437}, 56 (1994)
\bibitem{Fischer1999} P. Fischer: AJ \textbf{117}, 2024 (1999)

\bibitem{FT97} P. Fischer, J.~A. Tyson: AJ \textbf{114}, 14 (1997)

\bibitem{octupole} D. Goldberg, M. Natarajan: ApJ, submitted,
astro-ph/0107187 (2001)

\bibitem{Gray2001a} M. Gray, R. Ellis, J. Lewis, R. McMahon, 
A. Firth: MNRAS \textbf{325}, 111 (2001)

\bibitem{Gray2001} M. Gray, A. Taylor, K. Meisenheimer, S. Dye,
C. Wolf, E. Thommes: ApJ \textbf{568}, 141 (2002)

\bibitem{Grego_etal} L. Grego, J. Carlstrom, E. Reese, G. Holder,
W. Holzapfel, M. Joy, J. Mohr, S. Patel: ApJ \textbf{552}, 2 (2001)

\bibitem{Guimaraes} A. Guimaraes: MNRAS, submitted, astro-ph/0202507 (2002)

\bibitem{Gunn67} J. Gunn: ApJ \textbf{147}, 61 (1967)

\bibitem{HRH2000} A. Heavens, A. Refregier, C. Heymans:
MNRAS \textbf{319}, 649 (2000)

\bibitem{Hennawi} J. Hennawi, V. Narayanan, D. Spergel,
I. Dell'Antonio, V. Margoniner, J.~A. Tyson, D. Wittman: BAAS
\textbf{199}, 1608 (2001)

\bibitem{Hoekstra} H. Hoekstra: A\&A \textbf{370}, 743 (2001)

\bibitem{Hoekstra_etal2001} H. Hoekstra, M. Franx, K. Kuijken,
R.G. Carlberg, H.K.C. Yee, H. Lin, S.L. Morris, P.B. Hall,
D.R. Patton, M. Sawicki, G.D. Wirth: ApJL \textbf{548}, L5 (2001)

\bibitem{HYG2001} H. Hoekstra, H. Yee, M. Gladders: to appear in the
proceedings of the STScI 2001 spring symposium "Dark Universe",
astro-ph/0106388 (2001)

\bibitem{Hogg} D. Hogg: astro-ph/9905116 (1999)

\bibitem{Hogg1998} D. Hogg, \etal: AJ \textbf{115}, 1418 (1998)

\bibitem{Holder} G. Holder, J. Mohr, J. Carlstrom, A. Evrard,
E. Leitch: ApJ \textbf{544}, 629 (2000)

\bibitem{Hu1999} W. Hu: ApJL \textbf{522}, L21 (1999)

\bibitem{Hu2000} W. Hu: PRD \textbf{62}, 3007 (2000)


\bibitem{Hu2001a} W. Hu: ApJL \textbf{557}, L79 (2001)

\bibitem{Hu2001b} W. Hu: Phys. Rev. D, submitted,
astro-ph/0108090 (2001)

\bibitem{HW2000} W. Hu, M. White: ApJ \textbf{554}, 67 (2001)

\bibitem{Huterer} D. Huterer: ApJ, submitted, astro-ph/0106399 (2001)

\bibitem{Kaiser1992} N. Kaiser: ApJ \textbf{388}, 272 (1992)


\bibitem{Kaiser1998} N. Kaiser: ApJ \textbf{498}, 26 (1998)

\bibitem{Kaiser1999} N. Kaiser: in {\it Gravitational Lensing: Recent
Progress and Future Goals}, ASP Conference Proceedings, Vol.~237.~
Edited by Tereasa G.~Brainerd and Christopher S.~Kochanek.~ San
Francisco: Astronomical Society of the Pacific, ISBN: 1-58381-074-9,
p.269 (2001)

\bibitem{Kaiser2000} N. Kaiser: ApJ \textbf{537}, 555 (2000)

\bibitem{KS93} N. Kaiser, G. Squires: ApJ \textbf{404}, 441 (1993)

\bibitem{KSB} N. Kaiser, G. Squires, T. Broadhurst: ApJ \textbf{449},
460 (1995)

\bibitem{Kaiser_supercluster} N. Kaiser, G. Wilson, G. Luppino,
L. Kofman, I. Gioia, M. Metzger, H. Dahle: ApJ, \textbf{} submitted,
astro-ph/9809268 (1998)

\bibitem{KWL} N. Kaiser, G. Wilson, G. Luppino: ApJL, \textbf{} submitted,
astro-ph/0003338 (2000)

\bibitem{Keeton2001} C. Keeton: astro-ph/0102341 (2001)

\bibitem{KS2000} L. King, P. Schneider: A\&A \textbf{369}, 1 (2001)

\bibitem{KIS2001} A. Knebe, R. Islam, J. Silk: MNRAS \textbf{326}, 109 (2001)

\bibitem{Kristian67} J. Kristian: ApJ \textbf{147}, 864 (1967)
 
\bibitem{KS66} J. Kristian, R. Sachs: ApJ \textbf{143}, 379 (1966)

\bibitem{KS1999} G. Kruse, P. Schneider: MNRAS \textbf{302}, 821 (1999)

\bibitem{Kuijken} K. Kuijken: astro-ph/0007368 (2000)


\bibitem{LP2000} J. Lee, U. Pen: ApJ \textbf{555}, 106 (2000)

\bibitem{LP2001} J. Lee, U. Pen: ApJL \textbf{567}, L111 (2000)

\bibitem{LP86} R. Lynds, V. Petrosian: BAAS \textbf{18}, 1014 (1986)

\bibitem{MWK2001} J. Mackey, M. White, M. Kamionkowski:
MNRAS, \textbf{332}, 788 (2002)

\bibitem{Maoli} R. Maoli, L. van Waerbeke, Y. Mellier, P. Schneider, B. Jain,
F. Bernardeau, T. Erben, B. Fort: A\&A \textbf{368}, 766 (2001)

\bibitem{MS2000} C. Mayen, G. Soucail: A\&A \textbf{361}, 415 (2000)

\bibitem{Mellier} Y. Mellier: ARAA \textbf{37}, 127 (1999)

\bibitem{MWL2001} C. Metzler, M. White, C. Loken: ApJ \textbf{547}, 560 (2001)

\bibitem{Jordi1991a} J. Miralda-Escud\'{e}: ApJ \textbf{370}, 1 (1991)

\bibitem{Jordi1991b} J. Miralda-Escud\'{e}: ApJ \textbf{380}, 1 (1991)

\bibitem{MB95} J. Miralda-Escud\'{e}, A. Babul: ApJ \textbf{449}, 18 (1995)

\bibitem{Mirallesetal2002}  J.-M. Miralles, T. Erben, H. Haemmerle,
P. Schneider, R.A.E. Fosbury, W. Freudling, N. Pirzkal, B. Jain,
S.D.M. White: A\&A submitted, astro-ph/0202122 (2002)

\bibitem{MJ98} R. Moessner, B. Jain: MNRAS \textbf{294}, 291 (1998)

\bibitem{MortlockTurner2001}D. J. Mortlock, E. L. Turner:
MNRAS \textbf{327}, 552 (2001)

\bibitem{Mould} J. Mould, R. Blandford, J. Villumsen, T. Brainerd,
I. Smail, T. Small, W. Kells: MNRAS \textbf{271}, 31 (1994)

\bibitem{NFW95} J. Navarro, C. Frenk, S. White: MNRAS \textbf{275}, 720 (1995)
\bibitem{NFW97} J. Navarro, C. Frenk, S. White: ApJ \textbf{490}, 493 (1997)

\bibitem{2dF} J. Peacock {\it et al.}: Nature \textbf{410}, 169 (2001)

\bibitem{Peebles} P. J. E. Peebles: {\it Principles of Physical
Cosmology}, Princeton University Press (1993)

\bibitem{PLS} U. Pen, Li, U. Seljak: ApJL \textbf{543}, L107 (2000)

\bibitem{modes} U. Pen, L. van Waerbeke, Y. Mellier: ApJ \textbf{567},
31 (2002)

\bibitem{RB99} K. Reblinsky, M. Bartelmann: A\&A \textbf{345}, 1 (1999)

\bibitem{Rhodes} J. Rhodes, A. Refregier, E. Groth: ApJL \textbf{552}, 85 (2001)

\bibitem{RGS2000} J. Robinson, E. Gawiser, J. Silk: ApJ \textbf{532}, 1 (2000)

\bibitem{shapelets1} A. Refregier: MNRAS, \textbf{} submitted,
astro-ph/0105178 (2001)

\bibitem{shapelets2} A. Refregier, D. Bacon: MNRAS, \textbf{} submitted,
astro-ph/0105179 (2001)

%
\bibitem{Saslaw} W. Saslaw: {\it Gravitational Physics of Stellar and
Galactic Systems}, Cambridge University Press (1987)

\bibitem{Sato} J. Sato, M. Takada, Y. P. Jing, T. Futamase: ApJL
\textbf{551}, L5 (2001)

\bibitem{SS95} P. Schneider, S. Seitz: A\&A \textbf{294}, 411 (1995)

\bibitem{Schneider1996} P. Schneider: MNRAS \textbf{283}, 837 (1996)

\bibitem{Schneider_apmass} P. Schneider, L. van Waerbeke, B. Jain,
G. Kruse: MNRAS \textbf{296}, 873 (1998)

\bibitem{Schuecker} P. Schuecker, H. Bohringer, H. Reiprich,
L. Feretti: A\&A \textbf{378}, 408 (2001)

\bibitem{Sellwood} J. Sellwood, A. Kosowsky: to appear in {\it The
Dynamics, Structure \& History of Galaxies}, G. S. Da Costa \&
E. M. Sadler, eds, ASP Conference Series; astro-ph/0109555 (2002)

\bibitem{Sheldon} E. Sheldon, {\it et al.}: 
ApJ \textbf{554}, 881 (2001)

\bibitem{Smail97} I. Smail, R. Ellis, A. Dressler, W. Couch,
A. Oemler, R. Sharples, H. Butcher: ApJ \textbf{479}, 70 (1997)

\bibitem{Soucail87} G. Soucail, B. Fort, Y. Mellier, J. Picat: A\& A
\textbf{172}, L14 (1987)

\bibitem{SK96} G. Squires, N. Kaiser: ApJ \textbf{473}, 65 (1996)

\bibitem{Takada2000} M. Takada, T. Futamase: ApJ \textbf{546}, 620 (2001)

\bibitem{Taruya} A. Taruya, M. Takada, T. Hamana, I. Kayo,
T. Futamase: ApJ, submitted, astro-ph/0202090 (2002)

\bibitem{Tyson88} J.~A. Tyson: AJ \textbf{96}, 1 (1988)

\bibitem{CL0024} J.~A. Tyson, G. Kochanski, I. Dell'Antonio: ApJ
\textbf{498}, 107 (1998)

\bibitem{TysonSeitzer1988} J.~A. Tyson, P. Seitzer:
ApJ \textbf{335}, 552 (1988)

\bibitem{TWV90} J.~A. Tyson, R. Wenk, F. Valdes: ApJL \textbf{349}, L1 (1990)

\bibitem{LSST} J. A. Tyson, D. Wittman, J. R. P. Angel: 
to appear in
proceedings of the Dark Matter 2000 conference (Santa Monica, February
2000) to be published by Springer, astro-ph/0005381 (2000)

\bibitem{DLS} J.~A. Tyson, D. Wittman, I. Dell'Antonio, A. Becker,
V. Margoniner, DLS Team: BAAS \textbf{199}, 10113 (2001)

\bibitem{Umetsu2000} K. Umetsu, T. Futamase: ApJL \textbf{539}, L5 (2000)

\bibitem{UzanBernardeau2001} J.-P. Uzan, F. Bernardeau: Phys. Rev. D
64, 3004 (2001)

\bibitem{VJT83} F. Valdes, J. Jarvis, J.~A. Tyson: ApJ \textbf{271}, 431 (1983)
 
\bibitem{vW2000} L. van Waerbeke \etal: A\&A \textbf{358}, 30 (2000)

\bibitem{vW2001} L. van Waerbeke \etal: A\&A \textbf{374}, 757 (2001)

%
%
\bibitem{Virgo} http://www.mpa-garching.mpg.de/Virgo/virgoproject.html

\bibitem{WhiteKochanek2001} M. White, C. S. Kochanek: ApJ
\textbf{560}, 539 (2001)

\bibitem{White2001} M. White, L. van Waerbeke, J. Mackey: 
astro-ph/0111490 (2001)

\bibitem{WKL2001} G. Wilson, N. Kaiser, G. Luppino: ApJ \textbf{556}, 601
(2001)

\bibitem{Wittmanetal2002} D. Wittman, V. Margoniner, J.~A. Tyson, 
J. Cohen, I. Dell'Antonio: ApJL in prep (2002)

\bibitem{Wittman2001} D. Wittman, J.~A. Tyson, V. Margoniner,
J. Cohen, I. Dell'Antonio: ApJL \textbf{557}, L89 (2001)

\bibitem{NCS} D. Wittman, I. Dell'Antonio, J.~A. Tyson, G.
Bernstein, P. Fischer, D. Smith: in {\it Constructing the Universe with
Clusters of Galaxies}, IAP 2000 meeting, Paris, France, July 2000,
Florence Durret \& Daniel Gerbal (Eds.) (2000)

 
\bibitem{WTKDB} D. Wittman, J.~A. Tyson, D. Kirkman, I. Dell'Antonio, G.
Bernstein: Nature \textbf{405}, 143 (2000)

\bibitem{SDSS} D. York {\it et al.}: AJ \textbf{120}, 1579 (2000)

\bibitem{Zaroubi} S. Zaroubi, G. Squires, G. de Gasperis, A. Evrard,
Y. Hoffman, J. Silk: ApJ \textbf{561}, 600 (2001)


\end{thebibliography}
\end{document}